\documentclass[aps, pra, twocolumn, notitlepage, groupedaddress, 10pt, floatfix]{revtex4-2}
\usepackage[pdftex,plainpages=false,colorlinks=true,citecolor=blue,linkcolor=blue,urlcolor=blue,filecolor=green,bookmarksopen=true]{hyperref}
\usepackage{amsmath}
\usepackage[caption = false]{subfig}
\usepackage{graphicx,epstopdf}
\usepackage[english]{babel}
\usepackage{blindtext}
\usepackage{lipsum}
\usepackage{amsfonts}
\usepackage{amssymb}
\usepackage{bbm}
\usepackage{enumerate}
\usepackage{color}
\usepackage{latexsym}
\usepackage{times,txfonts}
\usepackage{physics}

\DeclareSymbolFont{CMletters}{OML}{cmm}{m}{it}
\DeclareMathSymbol{J}{\mathalpha}{CMletters}{`J}
\DeclareMathSymbol{j}{\mathalpha}{CMletters}{`j}
\DeclareMathSymbol{U}{\mathalpha}{CMletters}{`U}
\newcommand{\identity}{1\!\!1}

\begin{document}

\title{Heat-based circuits using quantum rectification}

\author{Kasper Poulsen}
\email{poulsen@phys.au.dk}
\affiliation{Department of Physics and Astronomy, Aarhus University, Ny munkegade 120, 8000 Aarhus C, Denmark}

\author{Nikolaj T. Zinner}
\email{zinner@phys.au.dk}
\affiliation{Department of Physics and Astronomy, Aarhus University, Ny munkegade 120, 8000 Aarhus C, Denmark}

\begin{abstract}
With increased power consumption of modern computer components, heat-based circuitry has become ever more relevant due to a lower power expense to process logic bits of information. In heat-based circuits, computations are performed by driving heat currents through a circuit using a temperature difference. Utilizing harmonic oscillators and three-level quantum rectifiers as base components, we study three different heat-based circuits: a series configuration of diodes, a parallel configuration of diodes, and a diode bridge rectifier. We demonstrate the required functionality of each circuit for use as heat-based analogues of standard electronic components. Furthermore, the diode bridge rectifier is found to give consistent sign of the output bias independent of the input bias thus rectifying the input. Our results prove the theoretical feasibility of combining heat current components into heat-based circuits. The three circuits should be realizable using several of the current quantum technology platforms.
\end{abstract}

\maketitle

\section{Introduction}

As miniturization of classical transistors become increasingly difficult, the effort to keep Moore's law alive has led to increased energy consumption of computer components \cite{5696765}. It is hoped that not only charge but also alternative information carriers such as magnetic (spin) \cite{zutic2004, wolf2001, PhysRevLett.126.077203} and thermal (phonon) \cite{RevModPhys.84.1045, ROBERTS2011648} currents can be leveraged in future technologies. Thermal currents are driven by subjecting a heat circuit to a temperature difference, and heat-based circuits could, therefore, be powered by excess heat making it an especially promising platform through its low energy consumption.

With the development of quantum technologies, the understanding of heat transport necessary for heat-based circuitry has increased substantially. This includes theoretical proposals of minimal engines \cite{PhysRevE.76.031105,PhysRevLett.125.166802,cavaliere2022non,PhysRevLett.126.120605}, entanglement engines \cite{Bohr_Brask_2015}, and various versions of Maxwell's demon \cite{PhysRevA.56.3374,PhysRevLett.118.260603,PhysRevE.105.044141}. Additionally, experiments have proven the feasibility of heat control in systems such as Maxwell's demon \cite{PhysRevLett.113.030601,PhysRevLett.121.030604}, diodes \cite{Senior2020}, and heat valves \cite{Ronzani2018,Gubaydullin2022}.

Functioning heat-based circuitry requires a toolbox of versatile components which can be used to control the heat flow inside the circuits. Many components, such as transistors \cite{PhysRevLett.116.200601,PhysRevB.101.184510,PhysRevE.98.022118} and diodes \cite{poulsen2022dark,PhysRevB.98.035414,D_az_2021,PhysRevApplied.15.054050,PhysRevE.89.062109,PhysRevB.73.205415,PRXQuantum.3.010323, PhysRevB.103.155434,Tesser_2022,PhysRevA.105.052605}, have been proposed. 
Most proposals study the component exposed directly to a temperature bias i.e. a transistor or diode coupled directly to thermal baths at either end. However, to achieve heat-based circuits, more complex combinations of multiple components are required. Many of these components rely on this proximity to the thermal baths, and therefore, they can not trivially be used in larger circuits. One popular circuit in classical electronics is the full wave bridge rectifier, which is a bridge consisting of four diodes \cite{patent}. The purpose of the full wave bridge rectifier is to convert an alternating current into a direct current such that the output voltage is of the same sign independent of the input voltage.

\begin{figure}[h]
\begin{center}\includegraphics[width=1. \linewidth, angle=0]{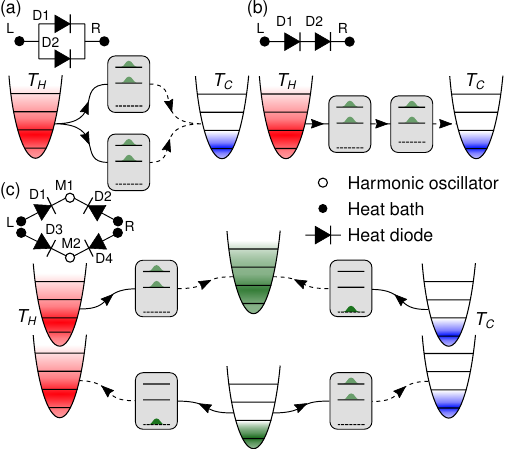}\end{center}
\caption{Schematic of the three different circuits studied: (a) parallel configuration of diodes, (b) series configuration of diodes, and (c) diode bridge rectifier. The circuits a shown both using a level diagram and more traditional circuit pictograms. The circuit pictograms include the numbering used throughout}
\label{figure1}
\end{figure}

Here, we study three different heat-based circuits as seen in Fig.~\ref{figure1}: a parallel configuration of diodes, a series configuration of diodes, and a diode bridge rectifier. All three circuits are comprised of harmonic oscillators and three-level quantum systems that can be operated as quantum rectifiers or diodes \cite{poulsen2022dark}. The general circuit setup is a heat source and a heat drain each coupled to a harmonic oscillator of equal frequency. These harmonic oscillators act as input and output filters setting the scale of the energy quanta of the heat current. The circuits are build up of components connecting the input and output harmonic oscillators. We show that the functionality of the components have added complexity when compared to a single component circuit. However, the functionality of the components is generally persistent across all three circuits, enabling their use as a heat-based alternative to their standard electronics counterparts.

\section{Results}

The Hamiltonian for all the circuits studied here takes the form
\begin{equation}
\hat{H} = \omega \Big( \hat{a}_L^\dag \hat{a}_L + \hat{a}_R^\dag \hat{a}_R \Big) + \hat{H}_\text{Circuit}, \label{eq:hamiltonian}
\end{equation}
where $\hat{a}_L$ and $\hat{a}_R$ are ladder operators for the left (L) and right (R) harmonic oscillators, respectively. $\hat{H}_{\text{Circuit}}$ is the Hamiltonian for the circuit connecting L and R. We are using units where $\hbar = k_B=1$. To drive a heat current through the circuit, L and R are coupled to thermal baths through the Lindblad master equation \cite{Lindblad1976, breuer2002theory}
\begin{equation}
\frac{d \hat{\rho}}{d t} = \mathcal{L}[\hat{\rho}] = -i [\hat{H}, \hat{\rho}] + \mathcal{D}_L[\hat{\rho}] + \mathcal{D}_R[\hat{\rho}], \label{eq:me}
\end{equation}
where $[\bullet, \bullet]$ is the commutator, $\mathcal{L}[\hat{\rho}]$ is the Lindblad superoperator, and $\mathcal{D}_{L(R)}[\hat{\rho}]$ is a dissipative term describing the action of the left (right) bath. Generally, thermal baths couple globally, driving transitions between the Floquet states of the system. However, the sensitivity of the Floquet spectrum towards different configurations of components makes it less likely for such configurations to function as intended. Instead, we assume that the baths act locally on the harmonic oscillators. This assumption is valid when the excitation energy $\omega$ is sufficiently large and the internal structure of the circuit becomes small, see Appendix \hyperref[app:validity]{A} for more information. Therefore, the dissipators become
\begin{equation}
\begin{aligned}
\mathcal{D}_{L(R)}[\hat{\rho}] &= \Gamma (n_{L(R)} + 1) \mathcal{M}[\hat{a}_{L(R)}, \hat{\rho}] + \Gamma n_{L(R)}   \mathcal{M}[\hat{a}_{L(R)}^\dag, \hat{\rho}], \label{eq:Dissipators}
\end{aligned}
\end{equation}
where $\mathcal{M}[\hat{A}, \hat{\rho}] = \hat{A} \hat{\rho} \hat{A}^\dag - \{\hat{A}^\dag \hat{A} , \hat{\rho}\}/2$, and $\{\bullet, \bullet\}$ is the the anti-commutator. $\Gamma$ is the coupling strength between the baths and harmonic oscillators and $n_{L(R)}~=~\big(e^{\omega/T_{L(R)}}-1\big)^{-1}$ is the mean number of excitations in the left (right) harmonic oscillator in the absence of the circuit. The harmonic oscillators act as filters of frequency $\omega$ between the baths and the circuit. After sufficient time the system will reach steady state. See Appendix \hyperref[methods]{B} for information on when the system is considered to have reached steady state. It is the properties of the steady state that we will focus on.
The circuit connecting L and R is comprised of harmonic oscillators and quantum rectifiers. The quantum rectifier used is a qutrit in the form of an anharmonic oscillator truncated to the three lowest states \cite{poulsen2022dark}, however, any three level system could be used. The Hamiltonian for the qutrit diode $\alpha$ connecting two harmonic oscillators $A$ and $B$, $\hat{H}^\alpha_{A\rightarrow B}$, is
\begin{equation}
\label{eq:qutrit_diode_hamiltonian}
\begin{aligned}
\hat{H}^{\alpha}_{A\rightarrow B} &+ \omega \left( \hat{a}_{A}^\dag \hat{a}_{A} + \hat{a}_{B}^\dag \hat{a}_{B} \right) = \\ 
& \hspace{1.cm} \omega \left( \hat{a}_{A}^\dag \hat{a}_{A} +  \hat{a}_\alpha^\dag \hat{a}_\alpha + \hat{a}_{B}^\dag \hat{a}_{B} \right) - \delta \omega_{\alpha} |{0_{\alpha}}\rangle \langle {0_{\alpha}}|   \\ &\hspace{1.cm} + J_{\alpha}(t) \Big(\hat{a}_A \hat{a}_{\alpha}^\dag + \hat{a}_A^\dag \hat{a}_\alpha \Big) + J \Big(\hat{a}_\alpha \hat{a}_B^\dag + \hat{a}_\alpha^\dag \hat{a}_B \Big), 
\end{aligned}
\end{equation}
where $\hat{a}_\alpha = |0_\alpha\rangle \langle 1_\alpha| + \sqrt{2} |1_\alpha\rangle \langle 2_\alpha|$. $|0_\alpha \rangle$, $|1_\alpha \rangle$, and $|2_\alpha \rangle$ are the three qutrit states. $\hat{a}_A$ and $\hat{a}_B$ are ladder operators of the harmonic oscillators connected by the qutrit. The harmonic oscillators are included in Eq.~\eqref{eq:qutrit_diode_hamiltonian} for clarity but excluded from the definition of $\hat{H}^{\alpha}_{A\rightarrow B}$ to avoid double counting. The two excited states of the qutrit have the same excitation energy as the harmonic oscillators namely $\omega$. The ground state is a dark state of B due to the anharmonicity $\delta \omega_\alpha$. The coupling to A is time-dependent with coupling strength $J_\alpha (t) = J + J'\cos (\delta \omega_{\alpha} t)$. 
\begin{figure}[]
\begin{center}\includegraphics[width=1.\linewidth, angle=0]{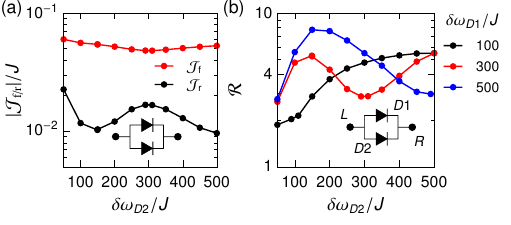}\end{center}
\caption{(a) Forward and reverse bias current as a function of the anharmonicity $\delta \omega_{D2}$ for a parallel configuration of diodes. (b) Rectification as a function of $\delta \omega_{D2}$ for a parallel configuration of diodes.
}
\label{figure2}
\end{figure}
In the limit $\omega \gg \delta \omega \gg J, J'$, the number of excitations exchanged between the left (right) harmonic oscillator and left (right) bath per unit of time is
\begin{equation}
\mathcal{J}_{L(R)} = \Gamma n_{L(R)} \langle \hat{a}_{L(R)} \hat{a}_{L(R)}^\dag \rangle_\text{ss} - \Gamma (n_{L(R)}+1) \langle \hat{a}_{L(R)}^\dag \hat{a}_{L(R)}\rangle_\text{ss}, \label{eq:current}
\end{equation}
where $\langle \bullet \rangle_\text{ss}= \text{tr} \left\{\bullet \hat{\rho}_\text{ss} \right\}$ is the steady state expectation value, and $\text{tr} \{\bullet\}$ is the trace over the entire Hilbert space. Due to the time-dependence of the Hamiltonian, the subscript ss implies a time average, e.g.,
\begin{align*}
\hat{\rho}_\mathrm{ss} = \frac{1}{T_\mathrm{av}} \int_{\tau}^{\tau+T_\mathrm{av}} \hat{\rho}(t) dt
\end{align*}
where $\tau$ is much larger than the system relaxation time and $T_\mathrm{av}=1000J$; See Appendix \hyperref[app:validity]{B} for more information.
$\mathcal{J}_{L(R)}$ is the excitation current whereas the heat current is given by $\omega \mathcal{J}_{L(R)}$. 
We will focus on the limit $\Gamma \gg J$ where L and R can be assumed to be in a thermal state due to the strong coupling to the baths. This is the Markovian limit for the two harmonic oscillators not to be confused with the original Markovian baths modeled through Eq.~\eqref{eq:Dissipators}. The two harmonic oscillators can be traced away thus lowering the degrees of freedom that has to be simulated, for further details see \cite{breuer2002theory,PhysRevE.105.044141,poulsen2021quantum}. We will rely on this approximation for the simulation of the system in Fig.~\ref{figure1}(c).
Unless otherwise stated, we use $\delta \omega_\alpha = 300J$, $J'=0.5J$, and $\Gamma = 20J$. 
The simplest circuit of one diode is
\begin{equation}
\hat{H}_\text{Circuit} = \hat{H}^{D1}_{L\rightarrow R},
\end{equation}
which has been studied in \cite{poulsen2022dark}. However, the functionality can be summarized in two parts. For $n_L > n_R$, 
excitations can propagate via transitions like
\begin{equation*}
\text{Forward:} \quad |1_L 1_{D1} 0_R \rangle \leftrightarrow |0_L 2_{D1} 0_R\rangle \leftrightarrow |0_L 1_{D1} 1_R \rangle.
\end{equation*}
For $n_L < n_R$, the qutrit is trapped in the dark state $|0_T\rangle$ through the transition
\begin{equation*}
\text{Reverse:} \quad |0_L 1_{D1} 0_R \rangle \leftrightarrow |1_L 0_{D1} 0_R\rangle \rightarrow |0_L 0_{D1} 0_R \rangle.
\end{equation*}
Since $|0_{D1}\rangle$ is a dark state of the right bath it will block transport. The simple excitation-based dark-state mechanism and the local bath couplings make this diode ideal for larger circuits. As long as excitations can propagate from the baths to either end of the diode, it is expected to work.
The parallel and series configuration of diodes as well as the heat-based diode bridge rectifier is designed to have this property. From classical electronics it is known that two diodes function like a single diode. Likewise, the diode bridge rectifier is known to yield a direction for the output bias which is independent of the input bias. The heat circuits are thus expected to work in a similar fashion.

\subsection{Parallel configuration of diodes}

First, we will study the rectification of a parallel configuration of qutrit diodes as seen in Fig.~\ref{figure1}(a). The Hamiltonian for the parallel configuration is
\begin{equation*}
\hat{H}_{\text{Circuit}} = \hat{H}^{D1}_{L\rightarrow R} + \hat{H}^{D2}_{L\rightarrow R}.
\end{equation*}
By forward bias, we denote the case where $n_L=0.1$ and $n_R=0$, and heat flows from left to right. By reverse bias, we denote the case where $n_L=0$ and $n_R=0.1$, and heat flows from right to left.
We define the forward bias excitation current to be $\mathcal{J}_\textrm{f} = -\mathcal{J}_R$, while the reverse bias excitation current is $\mathcal{J}_\textrm{r} = \mathcal{J}_L$. The quality of the diode is quantified using the rectification
\begin{equation*}
\mathcal{R} = -\frac{\mathcal{J}_\textrm{f}}{\mathcal{J}_\textrm{r}},
\end{equation*}
which tends to infinity for a perfect diode.
The forward and reverse bias currents can be seen in Fig.~\ref{figure2}(a). For $\delta \omega_{D1}=\delta \omega_{D2}$, the system is symmetric around the exchange of D1 and D2 which leads to an undefined steady state. Therefore, this value is avoided. The reverse bias current is clearly suppressed, as expected, resulting in modest rectification values. In Fig.~\ref{figure2}(b), the rectification for different values of $\delta \omega_{D1}$ and $\delta \omega_{D2}$ is plotted. A parallel configuration of diodes, therefore, results in a functioning diode. However, the rectification values are much smaller than the ones presented in \cite{poulsen2022dark}. Furthermore, the rectification drops around $\delta \omega_{D1}=\delta \omega_{D2}$ due to quantum correlation between the two diodes. Classically, an excitation would move from L to R, or the reverse, through one of the two diodes. This would result in a diode effect which is as good as the worst of the two diodes, in contrast to the rectification in Fig.~\ref{figure2}(b). Quantum mechanically, an excitation can be in a superposition of the two diodes resulting in markedly different behaviour around $\delta \omega_{D1} = \delta \omega_{D2}$ \cite{poulsen2021quantum}. In fact, the symmetry of parallel excitation paths can lead to additional dark states due to entanglement which might be useful for much larger rectification factors \cite{PhysRevA.105.052605}.

\begin{figure}[]
\begin{center}\includegraphics[width=1.\linewidth, angle=0]{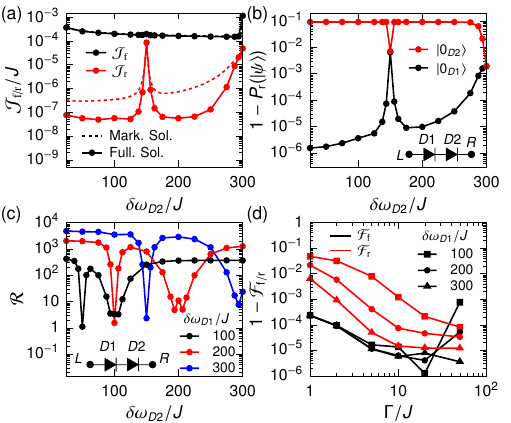}\end{center}
\caption{
(a) Forward and reverse bias current as a function of the anharmonicity $\delta \omega_{D2}$ for a series configuration of diodes. The dashed lines correspond to the approximates solution; See text. (b) Ground-state population of the two qutrits as a function of $\delta \omega_{D2}$ for a series configuration of diodes. (c) Rectification as a function of the anharmonicity $\delta \omega_{D2}$ for a series configuration of diodes. (d) Fidelity between the full and approximate solutions in forward and reverse bias as a function of the system bath coupling rate for $\delta \omega_{D2}=125J$.}
\label{figure3}
\end{figure}

\subsection{Series configuration of diodes}

Next, we study the rectification of a series configuration of qutrit diodes as seen in Fig.~\ref{figure1}(b). The most obvious series configuration consists of an extra harmonic oscillator added between two diodes. However, we will study a more compact version of just two diodes in series. The Hamiltonian of this series configuration is
\begin{equation*}
\hat{H}_{\text{Circuit}} = \hat{H}^{D1}_{L \rightarrow D2} + \hat{H}^{D2}_{D1 \rightarrow R} ,
\end{equation*}
where the diode Hamiltonians have to be modified to accommodate the compact circuit
\begin{equation*}
\begin{aligned}
\hat{H}^{D1}_{L \rightarrow D2} &= \omega \hat{a}_{D1}^\dag \hat{a}_{D1} - \delta \omega_{D1} |0_{D1}\rangle \langle 0_{D1}| + J_{D1}(t) \Big(\hat{a}_L \hat{a}_{D1}^\dag + \hat{a}_L^\dag \hat{a}_{D1} \Big), \\
\hat{H}^{D2}_{D1 \rightarrow R}  &= \omega \hat{a}_{D2}^\dag \hat{a}_{D2} - \delta \omega_{D2} |0_{D2}\rangle \langle 0_{D2}| +  \\ &\hspace{0.5cm} + J_{D2}(t) \Big(\hat{a}_{D1} \hat{a}_{D2}^\dag + \hat{a}_{D1}^\dag \hat{a}_{D2} \Big) + J \Big(\hat{a}_{D2} \hat{a}_R^\dag + \hat{a}_{D2}^\dag \hat{a}_R \Big). 
\end{aligned}
\end{equation*}
The coupling between the two qutrits is time-dependent to promote the decay of $D2$ to the ground state in reverse bias.
In Fig.~\ref{figure3}(a), the forward and reverse bias currents are plotted as a function of $\delta \omega_{D2}$. The reverse bias current has two peaks; a narrow peak at $\delta \omega_{D2} = \delta \omega_{D1}/2$ and a wider peak at $\delta \omega_{D2} = \delta \omega_{D1}$. To further investigate the drop in diode functionality at these two values, we plot the reverse-bias ground-state population of the two qutrits in Fig.~\ref{figure3}(b). We know from Ref.~\cite{poulsen2022dark} that a current is prevented due to the ground state being populated. Therefore, Fig.~\ref{figure3}(b) shows that the first qutrit (D1) is predominantly responsible for the large rectification. However, at the two special points $\delta \omega_{D2} = \delta \omega_{D1}/2$ and $\delta \omega_{D2} = \delta \omega_{D1}$, the populations are similar and the rectification is due to both qutrits. 
In Fig.~\ref{figure3}(c), the rectification is plotted for the series configuration of qutrit diodes. Rectification factors in the excess of $10^3$ are achieved, and the drops in rectification for $\delta \omega_{D2} = \delta \omega_{D1}/2$ and $\delta \omega_{D2} = \delta \omega_{D1}$ are still seen. So even though only one qutrit contributes to the rectification through ground state population, the overall rectification factors are still larger than for the parallel configuration. This is possible since the second qutrit (D2) can act as a filter making the effective interaction between the first qutrit (D1) and the right bath smaller. 

For comparison and for use later, an approximate solution is also found.
Since $\Gamma \gg J, J'$, we can trace L and R away resulting in a master equation for just the two qutrits. 
This is a Markov approximation for the two harmonic oscillators. The resulting system consists of two qutrits and two baths with Lorenzian spectral densities. These spectral densities are e.g. for the right bath
\begin{align*}
S_-^R(\omega') &= \frac{(1+n_R)J^2\Gamma}{(\omega'-\omega)^2 + \Gamma^2/4}, \\
S_+^R(\omega') &= \frac{n_R J^2\Gamma}{(\omega' +\omega)^2 + \Gamma^2/4},
\end{align*}
for decay and excitation, respectively. The baths, like before, drive transitions between Floquet states with rates, e.g., $S_-^R$ for decay due to the right bath; See Appendix \hyperref[app:validity]{A} for more information. Since $\Gamma \gg J,J'$, the Lorenzian spectral densities are approximately constant around the relevant transition frequencies $\pm \omega$, $\pm(\omega + \delta \omega_{D1})$, and $\pm(\omega + \delta \omega_{D2})$. 
Therefore, the local master equation is valid to a good approximation and becomes
\begin{equation}
\begin{aligned}
&\frac{d\hat{\rho}_{\mathrm{m}}}{dt} = \mathcal{L}[\hat{\rho}_\mathrm{m}] = -i[\hat{H}_{\text{Circuit}} - \hat{H}_{SB}, \hat{\rho}_{\mathrm{m}}] \\
&+ \sum_{q,k\in\mathcal{H}_{D1}} \Gamma_{q\rightarrow k}^{L}  \mathcal{M}[ |k\rangle \langle q|, \hat{\rho}_{\mathrm{m}}] 
+ \sum_{q,k\in\mathcal{H}_{D2}} \Gamma_{q\rightarrow k}^{R}  \mathcal{M}[ |k\rangle \langle q|, \hat{\rho}_{\mathrm{m}}] ,
\label{eq:me_mark2}
\end{aligned}
\end{equation}
where $\hat{\rho}_\mathrm{m}$ is the approximate density matrix for the two diodes, and $\hat{H}_{SB}$ is the new system bath coupling which is now included in the dissipative terms
\begin{equation*}
\hat{H}_{SB} = J_{D1}(t) \Big(\hat{a}_L \hat{a}_{D1}^\dag + \hat{a}_L^\dag \hat{a}_{D1} \Big) + J \Big(\hat{a}_{D2} \hat{a}_R^\dag + \hat{a}_{D2}^\dag \hat{a}_R \Big),
\end{equation*}
and the rates are
\begin{subequations}
\begin{alignat}{1}
\Gamma_{0_\alpha \rightarrow 1_\alpha}^L = \frac{n_L J'^2 }{\Gamma} + \frac{n_L J^2 \Gamma}{\delta \omega_\alpha^2 + \Gamma^2/4}, \,\, \Gamma_{1_\alpha \rightarrow 2_\alpha}^L = \frac{8n_L J^2}{\Gamma}, \hspace{1.1cm} \\
\Gamma_{1_\alpha \rightarrow 0_\alpha}^L = \frac{(1+n_L) J'^2 }{\Gamma} + \frac{(1+n_L ) J^2 \Gamma}{\delta \omega_{\alpha}^2 + \Gamma^2/4}, \,\, \Gamma_{2_\alpha \rightarrow 1_\alpha}^L = \frac{8(1+n_L) J^2}{\Gamma}, \\ 
\Gamma_{0_\alpha \rightarrow 1_\alpha}^R = \frac{n_R J^2 \Gamma}{\delta \omega_\alpha^2 + \Gamma^2/4}, \quad \Gamma_{1_\alpha \rightarrow 2_\alpha}^R = \frac{8n_R J^2}{\Gamma}, \hspace{1.6cm}\\
\Gamma_{1_\alpha \rightarrow 0_\alpha}^R = \frac{(1+n_R ) J^2 \Gamma}{\delta \omega_\alpha^2 + \Gamma^2/4}, \quad \Gamma_{2_\alpha \rightarrow 1_\alpha}^R = \frac{8(1+n_R) J^2}{\Gamma}, \hspace{1.1cm}
\end{alignat}
\end{subequations}
for $\alpha \in\{D1, D2\}$. All other rates are zero. These rates are a consequence of the Lorenzian spectral densities, and they are derived in greater detail in Ref.~\cite{poulsen2022dark}. The right bath rates $\Gamma^R_{q\rightarrow k}$ are similar to the left bath rates $\Gamma^L_{q\rightarrow k}$ without the term containing $J'$. Therefore, these rates will be useful whenever a diode is coupled to either the left or right harmonic oscillator. Additionally, the $J'$ term was not included for $\Gamma^L_{1_\alpha \rightarrow 2_\alpha}$ and $\Gamma^L_{2_\alpha \rightarrow 1_\alpha}$ since it is negligible for the parameter ranges studied here.
The transport is quantified using the excitation current which for this circuit becomes 
\begin{equation*}
\begin{aligned}
\mathcal{J}_\text{f} &= \langle 1_{D2} |\hat{\rho}_{\text{m,f,ss}}| 1_{D2} \rangle \Gamma_{1_{D2} \rightarrow 0_{D2}}^R +  \langle 2_{D2} |\hat{\rho}_{\text{m,f,ss}}| 2_{D2} \rangle \Gamma_{2_{D2} \rightarrow 1_{D2}}^R ,\\
\mathcal{J}_\text{r} &=- \left[ \langle 1_{D1} |\hat{\rho}_{\text{m,r,ss}}| 1_{D1} \rangle \Gamma_{1_{D1} \rightarrow 0_{D1}}^L +  \langle 2_{D1} |\hat{\rho}_{\text{m,r,ss}}| 2_{D1} \rangle \Gamma_{2_{D1} \rightarrow 1_{D1}}^L \right],
\end{aligned}
\end{equation*}
where $\hat{\rho}_\mathrm{m,f,ss}$ and $\hat{\rho}_\mathrm{m,r,ss}$ are the steady states in forward and reverse bias, respectively. 
The approximate currents are plotted as dashed lines in Fig.~\ref{figure3}(a). In forward bias, the approximate solution is very close to the full solution. However, in reverse bias, the two solutions deviate significantly. To investigate this discrepancy, we define the fidelity between the steady-state density matrices using the full theory and the approximate theory
\begin{align}
\mathcal{F}_{f/r} = F(\hat{\rho}^{D1,D2}_{\mathrm{f/r, ss}}, \hat{\rho}_{\mathrm{m, f/r, ss}}),
\end{align}
where $\hat{\rho}^{D1,D2}_{\mathrm{f/r,ss}} = \mathrm{tr}_{L,R} \{\hat{\rho}_{\mathrm{f/r, ss}} \}$, $\mathrm{tr}_{L,R}\{\cdot\}$ is the trace over the Hilbert space for the left and right harmonic oscillators, and $\hat{\rho}_{\mathrm{f/r, ss}}$ is the steady state density matrix in forward/reverse bias. $\hat{\rho}_\mathrm{m,f/r,ss}$ is the density matrix obtained using the approximate solution. The fidelity measure used here is 
\begin{align}
F(\hat{\rho}_1, \hat{\rho}_2) = \left[ \text{tr} \sqrt{\sqrt{\hat{\rho}_1} \hat{\rho}_2 \sqrt{\hat{\rho}_1}} \right]^2. \label{eq:fid}
\end{align}
In Fig.~\ref{figure3}(d), this fidelity between the two solutions is plotted as a function of $\Gamma$ for both forward and reverse bias. The Born-Markov approximation used to derive the approximate solution is valid when $\Gamma \gg J, J'$. This is supported by Fig.~\ref{figure3}(d), in which the fidelity increases for larger $\Gamma$. Importantly, the fidelity is close to one even in reverse bias. The current in reverse bias is very small and extremely sensitive to small variations in the excited state populations. Therefore, the approximate solution can be good for studying populations of appreciable size while populations of $\sim10^{-6}$ can not be trusted.

\begin{figure}[]
\begin{center}\includegraphics[width=1. \linewidth, angle=0]{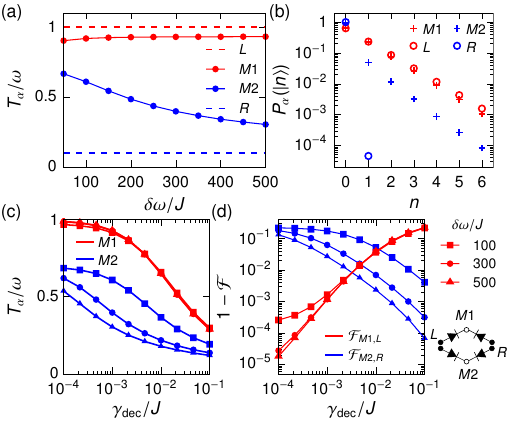}\end{center}
\caption{(a) Effective temperature of the harmonic oscillators as a function of the qutrit anharmonicity for the bridge rectifier. (b) Populations of the harmonic oscillators $P_\alpha (|n\rangle)$ for $\alpha = L,M1,M2,R$ for the bridge rectifier. (c) Effective temperature of the middle harmonic oscillators as a function of the decoherence rate $\gamma_\text{dec}$. (d) Fidelities between the states of the harmonic oscillators pairwise as a function of the decoherence rate $\gamma_\text{dec}$.}
\label{figure4}
\end{figure}

\subsection{Diode Bridge Rectifier}

The circuit for the bridge rectifier consists of four diodes and two additional harmonic oscillators as seen in Fig.~\ref{figure1}(c). If the qutrit diodes function as intended, the sign of the bias of the top and bottom harmonic oscillators will remain the same independent of the bias of the left and right harmonic oscillators. Since the circuit is horizontally symmetric, we can assume $T_L \geq T_R$. The two Hamiltonians for the top and bottom part of the bridge are
\begin{equation*}
\begin{aligned}
\mathrm{Top}:\, & \hat{H}_{\text{Circuit1}} =  \hat{H}^{D1}_{L1\rightarrow M1} + \hat{H}^{D2}_{R1\rightarrow M1} + \omega \hat{a}_{M1}^\dag \hat{a}_{M1}, \\
\mathrm{Bottom}:\, & \hat{H}_{\text{Circuit2}} = \hat{H}^{D3}_{M2 \rightarrow L2} + \hat{H}^{D4}_{M2\rightarrow R2} + \omega \hat{a}_{M2}^\dag \hat{a}_{M2} ,
\end{aligned}
\end{equation*}
where $\hat{a}_{M1}$ and $\hat{a}_{M1}$ are annihilation operators for the upper and lower harmonic oscillator, respectively. $\hat{a}_{L(R)1}$ and $\hat{a}_{L(R)2}$ are annihilation operators for the upper and lower left (right) harmonic oscillators. Due to the assumption $\Gamma \gg J,J'$, these are both traced away and have an identical density matrix. Therefore, we refer to both collectively as L(R). The two left and right harmonic oscillators are included to avoid correlations between the top and bottom circuit. We choose all anharmonicities the same, $\delta \omega$. The two effective master equations after the left and right harmonic oscillators have been traced away become
\begin{subequations}
\label{eq:me_mark3}
\begin{alignat}{1}
\mathrm{Top}:\, \mathcal{L}_1[\hat{\rho}] &= -i[\hat{H}_{\text{Circuit1}} - \hat{H}_{SB1}, \hat{\rho}] + \mathcal{K}^{D1}_{L1}[\hat{\rho}]  + \mathcal{K}^{D2}_{R1}[\hat{\rho}] \nonumber \\ 
&\hspace{-0.2cm} + \gamma_\text{dec} \sum_{\alpha_1} \Big( \mathcal{M}[\hat{a}_{\alpha_1}, \hat{\rho}] + \mathcal{M}[\hat{a}_{\alpha_1}^\dag \hat{a}_{\alpha_1}, \hat{\rho}] \Big), \\
\mathrm{Bottom}:\, \mathcal{L}_2[\hat{\rho}] &= -i[\hat{H}_{\text{Circuit2}} - \hat{H}_{SB2}, \hat{\rho}] + \mathcal{K}^{D3}_{L2}[\hat{\rho}] + \mathcal{K}^{D4}_{R2}[\hat{\rho}] \nonumber \\
&\hspace{-0.2cm} + \gamma_\text{dec} \sum_{\alpha_2} \Big( \mathcal{M}[\hat{a}_{\alpha_2}, \hat{\rho}] + \mathcal{M}[\hat{a}_{\alpha_2}^\dag \hat{a}_{\alpha_2}, \hat{\rho}] \Big) .
\end{alignat}
\end{subequations}
$\hat{H}_{SB1}$ and $\hat{H}_{SB2}$ are the system-bath couplings which are now included in the dissipative terms
\begin{align*}
\hat{H}_{SB1} &= J_{D1}(t) \Big(\hat{a}_{L1} \hat{a}_{D1}^\dag + \hat{a}_{L1}^\dag \hat{a}_{D1} \Big) + J_{D2}(t) \Big(\hat{a}_{D2} \hat{a}_{R1}^\dag + \hat{a}_{D2}^\dag \hat{a}_{R1} \Big) \nonumber \\
\hat{H}_{SB2} &= J \Big(\hat{a}_{L2} \hat{a}_{D3}^\dag + \hat{a}_{L2}^\dag \hat{a}_{D3} \Big) + J \Big(\hat{a}_{D4} \hat{a}_{R2}^\dag + \hat{a}_{D4}^\dag \hat{a}_{R2} \Big).
\end{align*}
The constant $\gamma_\text{dec}$ is the decoherence rate, which unless otherwise stated, is set to $\gamma_\text{dec}=10^{-3}J$. Decoherence due to both decay and dephasing is present in all experimental platforms. We keep the rate of decay and dephasing the same similar to platforms such as superconducting circuits \cite{Kim2023}. However, a large range of decoherence rates are studied to keep the theory general. The sums over $\alpha_1$ and $\alpha_2$ in Eq.~\eqref{eq:me_mark3} are carried out for $\alpha_1 \in \{D1, M1, D2\}$ and $\alpha_2 \in \{D3, M2, D4\}$. The four $\mathcal{K}$-terms describe the effective interaction of the baths with the four qutrits, e.g.,
$\mathcal{K}^{D1}_{L1}[\hat{\rho}] = \sum_{q,k\in\mathcal{H}_{D1}} \Gamma_{q\rightarrow k}^{L}  \mathcal{M}[ |k\rangle \langle q|, \hat{\rho}]$. To quantify the functionality of the bridge rectifier, we use the temperature of the two harmonic oscillators $M1$ and $M2$. Since the states of $M1$ and $M2$ are not necessarily a thermal state, we use the effective temperature
\begin{equation}
T_\alpha/\omega = \Big[ \ln\left(\langle \hat{n}_\alpha \rangle_{\text{ss}} + 1 \right) - \ln \langle \hat{n}_\alpha \rangle_{\text{ss}}\Big]^{-1},
\end{equation}
which is the temperature of a thermal state with the same occupation number $\langle \hat{n}_\alpha \rangle_{\text{ss}}$ as the harmonic oscillators. In Fig.~\ref{figure4}(a), the effective temperature of the four harmonic oscillators is plotted as a function of the diode quality parametrized through $\delta \omega$. This is compared to the temperatures of the left and right baths, $T_L = \omega$ and $T_R = 0.1\omega$, plotted with dashed lines. Under the above assumption of $T_L > T_R$ and since M1 and M2 are not connected, a perfect rectifier bridge would result in:
\[T_L \simeq T_{M1} \quad \text{and} \quad T_R \simeq T_{M2}.\]
If instead $T_L < T_R$, then L and R would be switched around in the above statement thus keeping the sign of the output bias the same. From Fig.~\ref{figure4}(a), we see that $T_{M1}$ quickly approaches $T_L$. The discrepancy between $T_{M1}$ and $T_{L}$ for large $\delta \omega$ is mainly due to decay through $\gamma_\text{dec}$. In Fig.~\ref{figure4}(b), the populations of the harmonic oscillator states are plotted. The populations for a thermal state with mean excitation number $\langle \hat{n}_\alpha \rangle_{\text{ss}}$ is
\begin{equation}
P_\alpha(n) = \frac{\langle \hat{n}_\alpha \rangle_{\text{ss}}^n}{\big[ 1+ \langle \hat{n}_\alpha \rangle_{\text{ss}} \big]^{n+1}} ,
\end{equation}
which is the population of L and R within the Markov approximation. Since Fig.~\ref{figure4}(b) is a log plot, thermal states will be points along a line. Harmonic oscillator M1 does indeed overlap with the thermal state of L. Harmonic oscillator M2 is clearly cold, but the state is not a thermal state. For M2 to get cold in the long time limit, diode D3 needs to close i.e. be driven into the ground state $|0_{D3}\rangle$. This can either be done by the cold right bath or through the decoherence rate $\gamma_{\text{dec}}$. To test the importance of $\gamma_{\text{dec}}$, we plot the effective temperatures as a function of $\gamma_{\text{dec}}$ for three different values of $\delta \omega$ in Fig.~\ref{figure4}(c). Clearly, decay through $\gamma_{\text{dec}}$ is important for M2 to be cold. However, for larger $\gamma_{\text{dec}}$, the effective temperature of M1 decreases due to decay. Since the states of M1 and M2 are not necessarily thermal, it is interesting to look at the fidelity between the relevant pairs of states. For this, we define the fidelities
\begin{align*}
\mathcal{F}_{M1,L} = F(\hat{\rho}^{M1}_{\mathrm{ss}}, \hat{\rho}^L_{\mathrm{ss}}), \\
\mathcal{F}_{M2,R} = F(\hat{\rho}^{M2}_{\mathrm{ss}}, \hat{\rho}^R_{\mathrm{ss}}),
\end{align*}
where the fidelity $F$ is defined in Eq.~\eqref{eq:fid}.
The density matrix for $\alpha \in \{L, M1, M2, R\}$ is $\hat{\rho}^\alpha_\mathrm{ss} = \text{tr}_{S \backslash \alpha}\{\hat{\rho}_\mathrm{ss}\}$ where the trace is over the entire system except $\alpha$. The fidelities between two pairs of states $\mathcal{F}_{M1,L}$ and $\mathcal{F}_{M2,R}$ are plotted in Fig.~\ref{figure4}(d). The fidelity between the state of M1 and L becomes smaller for larger $\gamma_\text{dec}$, while the state of M2 becomes closer to the state of R. So picking $\gamma_{\text{dec}}$ is clearly a balance between getting $T_L \simeq T_{M1}$ for small $\gamma_{\text{dec}}$ and $T_R \simeq T_{M2}$ for larger $\gamma_{\text{dec}}$.

\section{Conclusion}

We have studied three different heat-based circuits: a parallel configuration of diodes, a series configuration of diodes, and a diode bridge rectifier. The parallel configuration of diodes was found to result in an effective diode with modest rectification values smaller than those of a single diode. The series configuration of diodes was found to result in an effective diode with similar or larger rectification factors as compared to a single diode. Larger rectification factors were found for two qutrit diodes of different anharmonicities. Finally, it was found that a heat-based diode bridge rectifier can be constructed similar to the full wave bridge rectifier in electronics. One output harmonic oscillator reaches the temperature of the hot bath, and for large anharmonicities, the other output harmonic oscillator reaches the temperature of the cold bath. The cold output only got cold when decoherence was included, and the ideal decoherence rate $\gamma_\text{dec}/J$ was found to be $10^{-3}-10^{-2}$. This is within the capabilities of current quantum technology platforms such as superconducting circuits \cite{doi:10.1146/annurev-conmatphys-031119-050605, Ronzani2018, PhysRevApplied.10.054062}, and trapped ions \cite{Johanning_2009,HAFFNER2008155}. In superconducting circuits, the circuits can be implemented using transmons for the qutrits and microwave resonators for the harmonic oscillators. The harmonic oscillator correlation functions can be forced to decay through a resistance coupled to the resonator \cite{Ronzani2018}. Additionally, tunable couplers allow for a time-dependent coupling \cite{PhysRevApplied.10.054062}. All system parameters can be chosen much smaller than the excitation frequency, validating the use of the local master equation.

\section*{Acknowledgements}
The authors acknowledge funding from The Independent Research Fund Denmark DFF-FNU. The numerical results presented in this work were obtained at the Centre for Scientific Computing, Aarhus.

\begin{figure}[]
\begin{center}\includegraphics[width=1. \linewidth, angle=0]{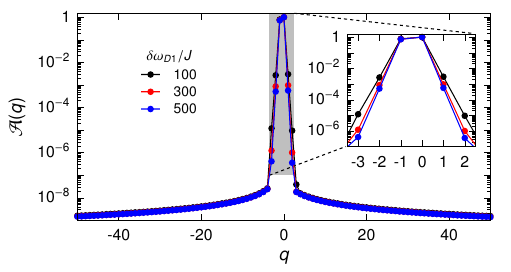}\end{center}
\caption{Frequency spectrum maximized over the Floquet states in the computational basis $\ket{\alpha_{L} \alpha_{D1} \alpha_{L}}$ for $\alpha \in \{0, 1, 2, ...\}$ as a function of the frequency $q\delta \omega_{D1}$. The numerical error is $\sim 10^{-7}$, and the inset shows all components that are not zero within the numerical error.}
\label{figure6}
\end{figure}

\section*{Appendix A: Validity of the local master equation.}
\label{app:validity}

The derivation of the master equation \eqref{eq:me} follows the structure of Ref.~\cite{breuer2002theory}. However, due to the time dependence of the Hamiltonian, essential points are taken from Refs.~\cite{PhysRevA.98.052129, PhysRevE.85.061126, PhysRevA.73.052311}. The total Hamiltonian for the system and bath is defined as
\begin{align*}
\hat{H}_{\mathrm{tot}}(t) = \hat{H}(t) \otimes \identity + \identity \otimes \hat{H}_B + \hat{H}_I(t). 
\end{align*}
where $\hat{H}(t)$ is the time-dependent system Hamiltonian from eq.~\eqref{eq:hamiltonian}, $\hat{H}_B$ is the bath Hamiltonian, and $\hat{H}_I(t)$ is the time-dependent interaction Hamiltonian between the system and bath. The system Hamiltonian is assumed periodic in time, i.e., $\hat{H}(t + T) = \hat{H}(t)$, with period $T$. The interaction Hamiltonian is written on the form
\begin{align*}
\hat{H}_I(t) = \sum_\alpha g_\alpha(t) \hat{A}_\alpha \otimes \hat{B}_\alpha.
\end{align*}
The time dependence of the interaction Hamiltonian enters through the complex functions $g_\alpha (t)$. 
First, the interaction Hamiltonian is moved into the interaction picture with respect to the rest of the Hamiltonian 
\begin{align*}
\hat{H}_{I,I}(t) = \hat{U}_\mathrm{SB}^\dag(t,0) \hat{H}_I \hat{U}_\mathrm{SB}(t,0)
\end{align*}
where the subscript $I$ is used to signify the interaction picture for operators that are already time dependent in the Schrödinger picture. The time evolution operator for the system and bath is
\begin{align*}
U_\mathrm{SB}(t_2, t_1) = U(t_2, t_1) \otimes e^{ -i \hat{H}_B (t_2-t_1) },
\end{align*}
where
\begin{align*}
U(t_2, t_1) = \mathcal{T} \exp \left( -i \int_{t_1}^{t_2} \hat{H}(s) ds \right),
\end{align*}
and $\mathcal{T}$ is the time ordering operator. The time evolution of the density matrix for the total system and bath setup is determined by the von Neumann equation
\begin{align*}
\frac{d }{dt} \hat{\rho}_{\mathrm{tot},I}(t) = -i [\hat{H}_{I,I}(t), \hat{\rho}_{\mathrm{tot},I}(t)].
\end{align*}
Following the method of Ref.~\cite{breuer2002theory}, this equation is integrated and put back into itself. 
\begin{align}
\frac{d \hat{\rho}_I}{ dt } &= \sum_{\alpha, \beta} \int_0^\infty ds \left[\hat{A}_\beta (t-s) \hat{\rho}_I (t) \hat{A}_\alpha^\dag (t) - \hat{A}_\alpha^\dag (t) \hat{A}_\beta (t-s) \hat{\rho}_I(t) \right] \nonumber \\
& \hspace{2cm} \times g_\beta(t-s) g_\alpha^*(t) \langle \hat{B}_\alpha^\dag (s) \hat{B}_\beta (0) \rangle_B + \mathrm{h.c.} \label{eq:RedField}
\end{align}
Additionally, the Born-Markov approximation was used and the bath was traced away using $\hat{\rho} = \tr_B \{ \hat{\rho}_\mathrm{tot} \}$. The Born approximation consists of writing $\hat{\rho}_\mathrm{tot}(t) \simeq \hat{\rho}(t) \otimes \hat{\rho}_B$, which is true when correlations between the bath and system are small. The Markov approximation is valid when the bath correlation functions $\langle \hat{B}_\alpha^\dag (s) \hat{B}_\beta (0) \rangle_B$ decay faster than the time it takes the system to change. Due to Floquet theory it is known that a set of states, called Floquet states, exist, which evolve as
\begin{align}
\hat{U}(t, 0) \ket{\phi_n (0) } = e^{-i\epsilon_n t} \ket{\phi_n (t)} \label{eq:floquet_states}
\end{align}
where $\epsilon_n$ are the Floquet energies and $\ket{\phi_n (t)} = \ket{\phi_n (t+T)}$. Setting $t=T$ above gives an eigenvalue problem where $\epsilon_n$ is only uniquely defined for $- \pi/T \leq \epsilon_n \leq \pi/T$. Since the Floquet states are periodic in time, they can be decomposed using a Fourier series
\begin{align}
\hat{U}(t,0) \ket{\phi_n (0)} = e^{-i\epsilon_n t} \sum_{q\in \mathbb{Z}} e^{-iq \Omega t} \ket{\phi_n (q)}, \label{eq:floquet_fourier}
\end{align}
where $\Omega = 2\pi/T$ and $q$ is a whole number. The Floquet states $ \ket{\phi_n (t)}$ form an orthonormal basis, while the states $\ket{\phi_n (q)}$ are not necessarily orthogonal or even normalized. Writing the system operators $\hat{A}_\alpha$ in the Floquet basis and using the above time evolution, the system operators can be decomposed into terms rotating with a single frequency
\begin{align}
\hat{A}_\alpha (t) 
&= \sum_{\Delta \epsilon}  \sum_{q\in \mathbb{Z}} e^{-i(\Delta \epsilon + \Omega q) t} \hat{A}_\alpha(\Delta \epsilon, q), \label{eq:A_floquet}
\end{align}
where
\begin{align}
\hat{A}_\alpha(\Delta \epsilon, q) = \sum_{\epsilon_m - \epsilon_n = \Delta \epsilon} \sum_{p\in \mathbb{Z}}  \bra{\phi_n (p)} \hat{A}_\alpha  \ket{\phi_m (p+q)} \ket{\phi_n} \! \bra{\phi_m}. \label{eq:A_floquet2}
\end{align}
We have used the simplified notation $\ket{\phi_n} = \ket{\phi_n(0)}$. Putting this back into Eq.~\eqref{eq:RedField} gives a general master equation for time dependent system and interaction Hamiltonians
\begin{align}
\frac{d \hat{\rho}_I}{dt} &= \sum_{q,q'} \sum_{\Delta \epsilon, \Delta \epsilon'} \sum_{\alpha, \beta} e^{i(\Delta \epsilon' + q' \Omega - \Delta \epsilon -q\Omega) t} \Gamma_{\alpha \beta} (\Delta \epsilon + q \Omega) \label{eq:floquet_master_equation} \\
& \hspace{-0.3cm} \left( \hat{A}_\beta (\Delta \epsilon, q) \hat{\rho}_I(t) \hat{A}_\alpha^\dag (\Delta \epsilon', q') - \hat{A}_\alpha^\dag (\Delta \epsilon', q') \hat{A}_\beta (\Delta \epsilon, q) \hat{\rho}_I(t) \right) + \mathrm{h.c.}, \nonumber
\end{align}
where
\begin{align*}
\Gamma_{\alpha \beta} (\Delta \epsilon) = \int_{0}^\infty ds \, e^{i \Delta \epsilon t} g_\beta (t-s) g_\alpha^*(t)  \langle \hat{B}^\dag_\alpha (s) \hat{B}_\beta (0) \rangle
\end{align*}
This master equation is essentially the same as the one derived Ref.~\cite{PhysRevA.73.052311} before the secular approximation and with a time-dependent interaction Hamiltonian. This master equation is complicated and difficult to use. However, for the systems discussed in the main text, it simplifies dramatically.

\subsection*{Single diode}

To see this, we will start with a single diode and generalize to multiple diodes later. The Hamiltonian for a single diode is
\begin{align*}
\hat{H} &= \hat{H}_0 + J_{D1}(t) \Big(\hat{a}_L \hat{a}_{D1}^\dag + \hat{a}_L^\dag \hat{a}_{D1} \Big) + J \Big(\hat{a}_{D1} \hat{a}_R^\dag + \hat{a}_{D1}^\dag \hat{a}_R \Big),\\
\hat{H}_0 &= \omega \left( \hat{a}_{L}^\dag \hat{a}_{L} + \hat{a}_{R}^\dag \hat{a}_{R} \right) +\omega \hat{a}_{D1}^\dag \hat{a}_{D1} - \delta \omega_{D1} |{0_{D1}}\rangle \langle {0_{D1}}| ,
\end{align*}
where everything except the first term in $\hat{H}_0$ corresponds to $\hat{H}_{\mathrm{Circuit}}$ from the main text. The time dependent coupling is $J_{D1} = J+ J' \cos (\delta \omega_{D1} t)$ as in the main text, which is periodic such that the above applies. First, the Floquet states are found by going into the interaction picture with respect to $\hat{H}_0$
\begin{align}
& e^{i\hat{H}_0 t} \hat{H} e^{-i\hat{H}_0 t}- \hat{H}_0 = \label{eq:diode_interaction_picture} \\
&\hspace{1.5cm} J_{D1}(t) \Big(\hat{a}_L \op{1_{D1}}{0_{D1}} e^{i\delta \omega_{D1} t} + \hat{a}_L^\dag \op{0_{D1}}{1_{D1}} e^{-i\delta \omega_{D1} t} \Big) \nonumber\\ 
& \hspace{1.5cm} + \sqrt{2} J_{D1}(t) \Big(\hat{a}_L \op{2_{D1}}{1_{D1}} + \hat{a}_L^\dag \op{1_{D1}}{2_{D1}} \Big) \nonumber \\
& \hspace{1.5cm} + J \Big(\op{0_{D1}}{1_{D1}} \hat{a}_R^\dag e^{-i\delta \omega_{D1} t} + \op{1_{D1}}{0_{D1}} \hat{a}_R e^{i\delta \omega_{D1} t} \Big) \nonumber\\
& \hspace{1.5cm} + \sqrt{2} J \Big(\op{1_{D1}}{2_{D1}} \hat{a}_R^\dag + \op{2_{D1}}{1_{D1}} \hat{a}_R \Big), \nonumber
\end{align}
Now, all the terms inducing transitions which do not obey energy conservation with respect to $\hat{H}_0$ contain time dependent phases. We temporarily perform the rotating wave approximation removing the terms rotating with frequency $\sim \delta \omega_{D1}$
\begin{align*} 
e^{i\hat{H}_0 t} \hat{H} e^{-i\hat{H}_0 t}- \hat{H}_0 &\simeq \hat{H}_{\mathrm{RWA}} \\
 &= \frac{J'}{2} \Big(\hat{a}_L \op{1_{D1}}{0_{D1}} + \hat{a}_L^\dag \op{0_{D1}}{1_{D1}} \Big) \\ 
& \hspace{0.5cm} + \sqrt{2} J \Big(\hat{a}_L \op{2_{D1}}{1_{D1}} + \hat{a}_L^\dag \op{1_{D1}}{2_{D1}} \Big) \\
& \hspace{0.5cm} + \sqrt{2} J \Big(\op{1_{D1}}{2_{D1}} \hat{a}_R^\dag + \op{2_{D1}}{1_{D1}} \hat{a}_R \Big),
\end{align*}
Let $\ket{\psi_n}$ be the eigenvector of $\hat{H}_\mathrm{RWA}$. Since $\delta \omega_{D1} \gg J, J'$, these states are, to a good approximation, also eigenstates of Eq.~\eqref{eq:diode_interaction_picture}. Going back out of the interaction picture these eigenstates become time-dependent Floquet states
\begin{align}
e^{i\epsilon_n t} \ket{\phi_n(t)} &\simeq e^{i\delta \omega_{D1} |0_{D1} \rangle \langle 0_{D1}|  t} \sum_{i \in \{0, 1, 2\}} \langle i_{D1} |\psi_n \rangle \ket{i_{D1}} \label{eq:floquet_states2} \\
&=  e^{i\delta \omega_{D1} t} \langle 0_{D1} |\psi_n\rangle \ket{0_{D1}}+ \sum_{i \in \{1, 2\}} \langle i_{D1} |\psi_n\rangle \ket{i_{D1}} \nonumber
\end{align}
Since the Hamiltonian is excitation preserving, all the eigenstates $\ket{\psi_n}$ have well defined particle numbers, and the first two terms of $\hat{H}_0$ only contribute with an irrelevant phase. The main takeaway from Eq.~\eqref{eq:floquet_states2} is that we can expect the Floquet states to have only two terms in the Fourier expansion \eqref{eq:floquet_fourier}. Instead of an infinite sum, $q$ should only take values of $q=-1$ and $q=0$. To verify this, we define the maximum of the norm of the Fourier coefficients
\begin{align*}
\mathcal{A}(q) = \mathrm{max}_{n, i_L, i_{D1}, i_R} \{| \langle i_L i_{D1} i_R |\phi_n (q)\rangle | \}
\end{align*}
The Floquet states are found numerically through Eq.~\eqref{eq:floquet_states}. Maximization is carried out over all the Floquet states through $n$, and all the coefficients in the basis $\{\ket{i_L i_{D1}  i_R}\}$. $\mathcal{A}(q)$ is plotted as a function of $q$ in Fig.~\ref{figure6} for different values of $\delta \omega_{D1}$. It is clear that the two components $q=-1$ and $q=0$ dominate the Floquet states. Other components do contribute a little, and if more precision is needed, $q=-2$ and $q=1$ could be kept. However, any other components are negligible. Likewise the sum in Eq.~\eqref{eq:A_floquet} can be carried out for $-2 \geq q \geq 2$ if needed. To a good approximation the states can, therefore, be written 
\begin{align*}
\ket{\phi_n(t)} \simeq e^{-i\epsilon_n t} \left( \ket{\phi_n(q=0)} + e^{i\delta \omega_{D1} t} \ket{\phi_n(q=-1)} \right)
\end{align*}
The state $\ket{\phi_n(q=0)}$ is a sum over all states with the qutrit in either the first or second excited states, and the state $\ket{\phi_n(q=-1)}$ is a sum over all states with the qutrit in the ground state. Therefore, the two components are orthogonal through the state of the qutrit. Importantly, for any bath operator $\hat{A}_\alpha$ which does not change the state of the qutrit, only $\hat{A}_\alpha (\Delta \epsilon, q=0)$ is non-zero. Any other component of $\hat{A}_\alpha(t)$ in Eq.~\eqref{eq:A_floquet2} is zero for all values of $\Delta \omega$. 
The final crucial property of the Floquet states comes from the conservation of the total number of excitations of the Hamiltonian. This means that, the states $\ket{\psi_n}$ has a well defined number of excitations, and the energy is, therefore, $\epsilon_n \sim k \omega$, where $k$ is a whole number. As a consequence, the difference in energies is likewise $\Delta \epsilon \sim k \omega$.

Having found these properties of the Floquet states, we return to the master equation with the system interaction operators $\hat{A}_L = \hat{a}_L + \hat{a}_L^\dag$ and $\hat{A}_R = \hat{a}_R + \hat{a}_R^\dag$. Additionally, we let $g_\alpha (t)= 1$. This couples the left and right harmonic oscillators to different baths, which can both add and subtract excitations from the respective harmonic oscillator. The correlation between the two baths is zero, and $\Omega = \delta \omega_{D1}$. Putting this in yields four times as many terms
\begin{align}
\frac{d \hat{\rho}_I}{dt} &= \sum_{\alpha\in\{L,R\}} \sum_{q,q'} \sum_{\Delta \epsilon, \Delta \epsilon'} e^{i(\Delta \epsilon' + q' \delta \omega_{D1} - \Delta \epsilon -q\delta \omega_{D1}) t} \Gamma_{\alpha \alpha} (\Delta \epsilon + q \delta \omega_{D1}) \nonumber \\
\times \bigg\{  &\hat{a}_\alpha (\Delta \epsilon, q) \hat{\rho}_I(t) \left[\hat{a}_\alpha (\Delta \epsilon', q')\right]^\dag - \left[\hat{a}_\alpha (\Delta \epsilon', q')\right]^\dag \hat{a}_\alpha (\Delta \epsilon, q) \hat{\rho}_I(t) \nonumber \\
 + &\hat{a}_\alpha (\Delta \epsilon, q) \hat{\rho}_I(t) \left[\hat{a}_\alpha^\dag (\Delta \epsilon', q')\right]^\dag - \left[\hat{a}_\alpha^\dag (\Delta \epsilon', q')\right]^\dag \hat{a}_\alpha (\Delta \epsilon, q) \hat{\rho}_I(t) \nonumber \\
+ &\hat{a}_\alpha^\dag (\Delta \epsilon, q) \hat{\rho}_I(t) \left[\hat{a}_\alpha (\Delta \epsilon', q')\right]^\dag - \left[\hat{a}_\alpha (\Delta \epsilon', q')\right]^\dag \hat{a}_\alpha^\dag (\Delta \epsilon, q) \hat{\rho}_I(t) \nonumber \\
 + &\hat{a}_\alpha^\dag (\Delta \epsilon, q) \hat{\rho}_I(t) \left[\hat{a}_\alpha^\dag (\Delta \epsilon', q')\right]^\dag - \left[\hat{a}_\alpha^\dag (\Delta \epsilon', q')\right]^\dag \hat{a}_\alpha^\dag (\Delta \epsilon, q) \hat{\rho}_I(t) \bigg\} \nonumber \\
 + &\mathrm{h.c.}, \label{eq:floquet_master_equation_diode}
\end{align}
where the hermition conjugate on the operators $\hat{A}_\alpha$ is kept separate for clarity. Since $\hat{a}_\alpha$ subtracts one excitation, the matrix elements $\langle \phi_n(p)| \hat{a}_\alpha | \phi_m(p+q) \rangle$ in Eq.~\eqref{eq:A_floquet2} is zero unless $\Delta \epsilon = \omega$. Similarly, $\langle \phi_n(p)| \hat{a}_\alpha^\dag | \phi_m(p+q) \rangle$ is zero unless $\Delta \epsilon = -\omega$. As mentioned above, they are both close to zero for $q\neq 0$. The first and fourth term will only have terms where $|\Delta \epsilon - \Delta \epsilon' | \sim 0$. Therefore, the term with $q'=0$ has a slowly rotating phase and contribute significantly to the dynamics. Since $\Delta \epsilon \sim \omega$, we can approximate $\Gamma_{\alpha \alpha} (\Delta \epsilon ) \simeq \Gamma_{\alpha \alpha} (\omega)$ for the first term, and $\Gamma_{\alpha \alpha} (\Delta \epsilon ) \simeq \Gamma_{\alpha \alpha} (-\omega)$ for the fourth term. 
Note that allowing terms with $q\neq 0$ in the sum for $\hat{A}_\alpha (\Delta \epsilon, q)$ can easily be done. If for example the $q=0$ and $q=-1$ terms in the approximate Floquet states are not assumed to be orthogonal, the following is still valid if $\Gamma_{\alpha \alpha} (\pm \omega + q \delta \omega_{D1}) \simeq \Gamma_{\alpha \alpha} (\pm \omega)$ where $q\leq q_{\mathrm{cutoff}}$. The cutoff $q_{\mathrm{cutoff}}$ can be picked to include as many terms as needed. From Fig.~\ref{figure6}, we see that the importance of the $q$th term in the actual Floquet states drops very sharply, and therefore, $q_{\mathrm{cutoff}}$ does not have to be very large.
The second and third terms of Eq.~ \eqref{eq:floquet_master_equation_diode} will only have terms where $|\Delta \epsilon - \Delta \epsilon' | \sim 2\omega$. These terms are quickly rotating and can be thrown away through a rotating wave approximation. Putting all of this together, we get
\begin{align*}
\frac{d \hat{\rho}_I}{dt} &= \sum_{\alpha\in\{L,R\}}  \Gamma_{\alpha \alpha} (\omega) \sum_{q,q'} \sum_{\Delta \epsilon, \Delta \epsilon'}  e^{i(\Delta \epsilon' + q' \delta \omega_{D1} - \Delta \epsilon -q\delta \omega_{D1}) t} \\
\times \bigg\{  &\hat{a}_\alpha (\Delta \epsilon, q) \hat{\rho}_I(t) \hat{a}_\alpha^\dag (\Delta \epsilon', q') - \hat{a}_\alpha^\dag (\Delta \epsilon', q') \hat{a}_\alpha (\Delta \epsilon, q) \hat{\rho}_I(t) \bigg\} \\
&+ \sum_{\alpha\in\{L,R\}} \Gamma_{\alpha \alpha} (-\omega) \sum_{q,q'} \sum_{\Delta \epsilon, \Delta \epsilon'} e^{i(\Delta \epsilon' + q' \delta \omega_{D1} - \Delta \epsilon -q\delta \omega_{D1}) t} \\
 \times \bigg\{ &\hat{a}_\alpha^\dag (\Delta \epsilon, q) \hat{\rho}_I(t) \hat{a}_\alpha (\Delta \epsilon', q') - \hat{a}_\alpha (\Delta \epsilon', q') \hat{a}_\alpha^\dag (\Delta \epsilon, q) \hat{\rho}_I(t) \bigg\} \\
 + &\mathrm{h.c.},
\end{align*}
Next, we move back into the Schrödinger picture and write $\Gamma_{\alpha \alpha}(\Delta \epsilon) = \gamma_\alpha(\Delta \epsilon)/2 + iS_\alpha (\Delta \epsilon)$, where $\gamma_\alpha$ is the real part of $\Gamma_{\alpha \alpha}$, and $S_\alpha$ is the imaginary part
\begin{align*}
\frac{d \hat{\rho}}{dt} &= - i[\hat{H}+\hat{H}_{LS}, \hat{\rho}] + \sum_{\alpha\in\{L,R\}}  \gamma_{\alpha} (\omega) \Big( \hat{a}_\alpha \hat{\rho}(t) \hat{a}_\alpha^\dag  - \{\hat{a}_\alpha^\dag \hat{a}_\alpha, \hat{\rho}(t) \} \Big) \\
&\hspace{1cm} + \sum_{\alpha\in\{L,R\}} \gamma_{\alpha} (-\omega) \Big( \hat{a}_\alpha^\dag \hat{\rho}(t) \hat{a}_\alpha - \{\hat{a}_\alpha \hat{a}_\alpha^\dag, \hat{\rho}(t)\} \Big) ,
\end{align*}
where $\hat{H}_{LS}$ is a Lamb shift given by
\begin{align*}
\hat{H}_{LS} = \sum_{\alpha\in \{L,R\} } \Big[ S_\alpha (\omega) + S_\alpha (-\omega) \Big] \hat{a}_\alpha^\dag \hat{a}_\alpha.
\end{align*}
This Lamb shift perturbs the frequency of the left and right oscillator, which we will assume is compensated for in the original frequency. Therefore, we can ignore $\hat{H}_{LS}$ without loss of generality. Finally, the decay and excitation rates are redefined using the notation in the main text
\begin{align*}
\gamma_\alpha (\omega) &= \Gamma (n_\alpha +1),\\
\gamma_\alpha (-\omega) &= \Gamma n_\alpha,
\end{align*}
resulting in the master equation from the main text.

\subsection*{Multiple diodes}

The approach for a single diode can easily be generalized to multiple diodes. For two diodes $D1$ and $D2$, the Hamiltonian contains two frequencies $\delta \omega_{D1}$ and $\delta \omega_{D2}$. We assume that $\delta \omega_{D1}$ and $\delta \omega_{D2}$ are rational numbers i.e., $\delta \omega_{D1} = n/m$ and $\delta \omega_{D2} = k/q$ for $n,m,k,q\in \mathbb{Z}$. This assumption can be made to arbitrary precision. Then the Hamiltonian has period
\begin{align*}
T = \frac{2 \pi}{\delta \omega_{D1}} nq = \frac{2 \pi}{\delta \omega_{D2}} mk,
\end{align*}
such that the master equation derived through Floquet theory still applies. The total Hamiltonian can again be moved into the interaction picture with respect to the non-interaction part of the Hamiltonian. For the parallel configuration of diodes the Hamiltonian is
\begin{align*}
\hat{H} &= \hat{H}_0 + \sum_{\alpha \in \{D1, D2 \}} \left\{ J_{\alpha}(t) \Big(\hat{a}_L \hat{a}_{\alpha}^\dag + \hat{a}_L^\dag \hat{a}_{\alpha} \Big) + J \Big(\hat{a}_{\alpha} \hat{a}_R^\dag + \hat{a}_{\alpha}^\dag \hat{a}_R \Big) \right\},\\
\hat{H}_0 &= \omega \left( \hat{a}_{L}^\dag \hat{a}_{L} + \hat{a}_{R}^\dag \hat{a}_{R} \right) +  \sum_{\alpha \in \{D1, D2 \}} \left\{ \omega \hat{a}_{\alpha}^\dag \hat{a}_{\alpha} - \delta \omega_{\alpha} |{0_{\alpha}}\rangle \langle {0_{\alpha}}| \right\},
\end{align*}
where $\hat{H_0}$ is the non-interaction part of the Hamiltonian. The eigenstates $\ket{\psi_n}$ can then be found after the rotating wave approximation. The Floquet states then becomes
\begin{align*}
\ket{\phi_n(t)} &\simeq \left( e^{i\delta \omega_{D1} t} |0_{D1}\rangle \langle 0_{D1}| + |1_{D1}\rangle \langle 1_{D1}| + |2_{D1}\rangle \langle 0_{D1}| \right) \\
& \hspace{0.7cm} \times \left( e^{i\delta \omega_{D2} t} |0_{D2}\rangle \langle 0_{D2}| + |1_{D2}\rangle \langle 1_{D2}|+ |2_{D2}\rangle \langle 2_{D2}| \right) \ket{\psi_n}
\end{align*}
Similar to the case of a single diode, the phase depends on the state of the two qutrits. If the qutrit $D1$ is in the ground state a phase with angular frequency $\delta \omega_{D1}$ is added, and similar for $D2$. The bath operator can only add or subtract one excitation at a time. Therefore, $\hat{A}_\alpha (t)$ does not contain terms rotating with frequency $\delta \omega_{D1} + \delta \omega_{D2}$ even though it shows up in the Floquet states. In fact the bath operators cannot change the qutrit state at all and $q=0$ dominates. Once again, more precision can be included by assuming $\Gamma_{\alpha \alpha}$ to be flat on the scale of $\delta \omega_{D1}$ and $\delta \omega_{D2}$ around $\pm \omega$. With this assumption, the derivation follows the single diode section. $\Gamma_{\alpha \alpha}$ can be pulled outside the sum in Eq.~\eqref{eq:floquet_master_equation}, and the bath operators are moved back into the Schrödinger picture. The resulting master equation is the one used in the main text. 

This approach can be generalized to any number of diodes. The overall period for the Floquet theory gets larger and larger, however, the theory remains valid.

\begin{figure}[]
\begin{center}\includegraphics[width=1. \linewidth, angle=0]{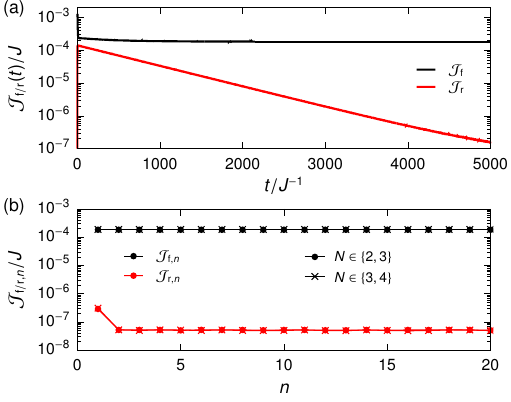}\end{center}
\caption{(a) Forward and reverse bias currents as a function of time for the series configuration of diodes. (b) Average currents in both forward and reverse bias as a function of the integer $n$ for a different number of harmonic oscillator levels $N$. For all plots $\delta \omega_{D1}=300J$, $\delta \omega_{D2}=125J$, and $n_{L(R)} \in \{0, 0.1\}$.}
\label{figure5}
\end{figure}

\section*{APPENDIX B: NUMERICAL METHODS AND CONVERGENCE}
\label{methods}

In the main text, we studied the properties of the steady-state. However, due to the time-dependent Hamiltonian the steady state will not obey $\partial_t \hat{\rho} = 0$, instead the density matrix is evolved until some expectation value converges. The harmonic oscillators are truncated to $N=3$ states for the hot harmonic oscillator in the parallel and series configuration of diodes, $N=2$ for the cold harmonic oscillator in the parallel and series configuration of diodes, and $N=8$ for the two harmonic oscillators in the diode bridge. This choice results in a population of $<10^{-3}$ for all truncated states in the Harmonic oscillator thermal state. The upper part of the bridge rectifier has a time-independent Hamiltonian, and the steady state can be solved through diagonalization. In Fig.~\ref{figure5}(a), the forward and reverse bias currents are plotted for the series configuration of diodes. To unsure that the current is allowed to converge, we use the average current
\begin{equation*}
\mathcal{J}_{\text{f/r}, n} = \frac{1}{T_{\text{av}}} \int_{nT -T_{\text{av}}}^{nT} \mathcal{J}_{\text{f/r}}(t) dt,
\end{equation*}
where $n$ determines for how long the density matrix is evolved, and we have picked $T = 5000J^{-1}$ and $T_{\text{av}} = 1000J^{-1}$. In Fig.~\ref{figure5}(b), the average current $\mathcal{J}_{\text{f/r}, n}$ is plotted as a function of $n$. We say that the current is converged when 
\begin{equation*}
\frac{\mathcal{J}_{\text{f/r}, n} - \mathcal{J}_{\text{f/r}, n-1}}{\mathcal{J}_{\text{f/r}, n-1}} < 10^{-4},
\end{equation*}
or after a time evolution corresponding to $n = n_{\mathrm{max}}$ and we set $\mathcal{J}_{\text{f/r}}=\mathcal{J}_{\text{f/r}, n}$. For the data presented in Fig.~\ref{figure2}, \ref{figure3}(a)-(c), and \ref{figure4}, we set $n_{\mathrm{max}} = 10$, while for Fig.~\ref{figure3}(d), we set $n_{\mathrm{max}} = 30$. In Fig.~\ref{figure4}, the convergence of $\langle \hat{n}_{M2} \rangle_{\mathrm{ss}}$ was used for the bottom half of the circuit. All steady state quantities with subscript ss have been averages over $T_\mathrm{av}$, whereas all other quantities are derived from these, e.g.,
\begin{align*}
\hat{\rho}_{\mathrm{ss}} = \frac{1}{T_{\text{av}}} \int_{n_\mathrm{conv} T -T_{\text{av}}}^{n_\mathrm{conv} T} \hat{\rho} (t) dt,
\end{align*}
where $n_\mathrm{conv}$ is the converged $n$ from above.  Finally, the currents for one additional level per harmonic oscillator are plotted Fig.~\ref{figure5}(b). The additional level has little effect on the currents.

\newpage


\begin{thebibliography}{44}%
\makeatletter
\providecommand \@ifxundefined [1]{%
 \@ifx{#1\undefined}
}%
\providecommand \@ifnum [1]{%
 \ifnum #1\expandafter \@firstoftwo
 \else \expandafter \@secondoftwo
 \fi
}%
\providecommand \@ifx [1]{%
 \ifx #1\expandafter \@firstoftwo
 \else \expandafter \@secondoftwo
 \fi
}%
\providecommand \natexlab [1]{#1}%
\providecommand \enquote  [1]{``#1''}%
\providecommand \bibnamefont  [1]{#1}%
\providecommand \bibfnamefont [1]{#1}%
\providecommand \citenamefont [1]{#1}%
\providecommand \href@noop [0]{\@secondoftwo}%
\providecommand \href [0]{\begingroup \@sanitize@url \@href}%
\providecommand \@href[1]{\@@startlink{#1}\@@href}%
\providecommand \@@href[1]{\endgroup#1\@@endlink}%
\providecommand \@sanitize@url [0]{\catcode `\\12\catcode `\$12\catcode
  `\&12\catcode `\#12\catcode `\^12\catcode `\_12\catcode `\%12\relax}%
\providecommand \@@startlink[1]{}%
\providecommand \@@endlink[0]{}%
\providecommand \url  [0]{\begingroup\@sanitize@url \@url }%
\providecommand \@url [1]{\endgroup\@href {#1}{\urlprefix }}%
\providecommand \urlprefix  [0]{URL }%
\providecommand \Eprint [0]{\href }%
\providecommand \doibase [0]{https://doi.org/}%
\providecommand \selectlanguage [0]{\@gobble}%
\providecommand \bibinfo  [0]{\@secondoftwo}%
\providecommand \bibfield  [0]{\@secondoftwo}%
\providecommand \translation [1]{[#1]}%
\providecommand \BibitemOpen [0]{}%
\providecommand \bibitemStop [0]{}%
\providecommand \bibitemNoStop [0]{.\EOS\space}%
\providecommand \EOS [0]{\spacefactor3000\relax}%
\providecommand \BibitemShut  [1]{\csname bibitem#1\endcsname}%
\let\auto@bib@innerbib\@empty
\bibitem [{\citenamefont {Mack}(2011)}]{5696765}%
  \BibitemOpen
  \bibfield  {author} {\bibinfo {author} {\bibfnamefont {C.~A.}\ \bibnamefont
  {Mack}},\ }\bibfield  {title} {\bibinfo {title} {Fifty years of moore's
  law},\ }\href {https://doi.org/10.1109/TSM.2010.2096437} {\bibfield
  {journal} {\bibinfo  {journal} {IEEE Trans. Semicond. Manuf.}\ }\textbf
  {\bibinfo {volume} {24}},\ \bibinfo {pages} {202} (\bibinfo {year}
  {2011})}\BibitemShut {NoStop}%
\bibitem [{\citenamefont {\ifmmode \check{Z}\else
  \v{Z}\fi{}uti\ifmmode~\acute{c}\else \'{c}\fi{}}\ \emph
  {et~al.}(2004)\citenamefont {\ifmmode \check{Z}\else
  \v{Z}\fi{}uti\ifmmode~\acute{c}\else \'{c}\fi{}}, \citenamefont {Fabian},\
  and\ \citenamefont {Das~Sarma}}]{zutic2004}%
  \BibitemOpen
  \bibfield  {author} {\bibinfo {author} {\bibfnamefont {I.}~\bibnamefont
  {\ifmmode \check{Z}\else \v{Z}\fi{}uti\ifmmode~\acute{c}\else \'{c}\fi{}}},
  \bibinfo {author} {\bibfnamefont {J.}~\bibnamefont {Fabian}},\ and\ \bibinfo
  {author} {\bibfnamefont {S.}~\bibnamefont {Das~Sarma}},\ }\bibfield  {title}
  {\bibinfo {title} {Spintronics: Fundamentals and applications},\ }\href
  {https://doi.org/10.1103/RevModPhys.76.323} {\bibfield  {journal} {\bibinfo
  {journal} {Rev. Mod. Phys.}\ }\textbf {\bibinfo {volume} {76}},\ \bibinfo
  {pages} {323} (\bibinfo {year} {2004})}\BibitemShut {NoStop}%
\bibitem [{\citenamefont {Wolf}\ \emph {et~al.}(2001)\citenamefont {Wolf},
  \citenamefont {Awschalom}, \citenamefont {Buhrman}, \citenamefont {Daughton},
  \citenamefont {Von~Molnar}, \citenamefont {Roukes}, \citenamefont
  {Chtchelkanova},\ and\ \citenamefont {Treger}}]{wolf2001}%
  \BibitemOpen
  \bibfield  {author} {\bibinfo {author} {\bibfnamefont {S.}~\bibnamefont
  {Wolf}}, \bibinfo {author} {\bibfnamefont {D.}~\bibnamefont {Awschalom}},
  \bibinfo {author} {\bibfnamefont {R.}~\bibnamefont {Buhrman}}, \bibinfo
  {author} {\bibfnamefont {J.}~\bibnamefont {Daughton}}, \bibinfo {author}
  {\bibfnamefont {S.}~\bibnamefont {Von~Molnar}}, \bibinfo {author}
  {\bibfnamefont {M.}~\bibnamefont {Roukes}}, \bibinfo {author} {\bibfnamefont
  {A.~Y.}\ \bibnamefont {Chtchelkanova}},\ and\ \bibinfo {author}
  {\bibfnamefont {D.}~\bibnamefont {Treger}},\ }\bibfield  {title} {\bibinfo
  {title} {Spintronics: a spin-based electronics vision for the future},\
  }\href {https://www.science.org/doi/abs/10.1126/science.1065389} {\bibfield
  {journal} {\bibinfo  {journal} {Science}\ }\textbf {\bibinfo {volume}
  {294}},\ \bibinfo {pages} {1488} (\bibinfo {year} {2001})}\BibitemShut
  {NoStop}%
\bibitem [{\citenamefont {Poulsen}\ and\ \citenamefont
  {Zinner}(2021)}]{PhysRevLett.126.077203}%
  \BibitemOpen
  \bibfield  {author} {\bibinfo {author} {\bibfnamefont {K.}~\bibnamefont
  {Poulsen}}\ and\ \bibinfo {author} {\bibfnamefont {N.~T.}\ \bibnamefont
  {Zinner}},\ }\bibfield  {title} {\bibinfo {title} {Giant magnetoresistance in
  boundary-driven spin chains},\ }\href
  {https://doi.org/10.1103/PhysRevLett.126.077203} {\bibfield  {journal}
  {\bibinfo  {journal} {Phys. Rev. Lett.}\ }\textbf {\bibinfo {volume} {126}},\
  \bibinfo {pages} {077203} (\bibinfo {year} {2021})}\BibitemShut {NoStop}%
\bibitem [{\citenamefont {Li}\ \emph {et~al.}(2012)\citenamefont {Li},
  \citenamefont {Ren}, \citenamefont {Wang}, \citenamefont {Zhang},
  \citenamefont {H\"anggi},\ and\ \citenamefont {Li}}]{RevModPhys.84.1045}%
  \BibitemOpen
  \bibfield  {author} {\bibinfo {author} {\bibfnamefont {N.}~\bibnamefont
  {Li}}, \bibinfo {author} {\bibfnamefont {J.}~\bibnamefont {Ren}}, \bibinfo
  {author} {\bibfnamefont {L.}~\bibnamefont {Wang}}, \bibinfo {author}
  {\bibfnamefont {G.}~\bibnamefont {Zhang}}, \bibinfo {author} {\bibfnamefont
  {P.}~\bibnamefont {H\"anggi}},\ and\ \bibinfo {author} {\bibfnamefont
  {B.}~\bibnamefont {Li}},\ }\bibfield  {title} {\bibinfo {title} {Colloquium:
  Phononics: Manipulating heat flow with electronic analogs and beyond},\
  }\href {https://doi.org/10.1103/RevModPhys.84.1045} {\bibfield  {journal}
  {\bibinfo  {journal} {Rev. Mod. Phys.}\ }\textbf {\bibinfo {volume} {84}},\
  \bibinfo {pages} {1045} (\bibinfo {year} {2012})}\BibitemShut {NoStop}%
\bibitem [{\citenamefont {Roberts}\ and\ \citenamefont
  {Walker}(2011)}]{ROBERTS2011648}%
  \BibitemOpen
  \bibfield  {author} {\bibinfo {author} {\bibfnamefont {N.}~\bibnamefont
  {Roberts}}\ and\ \bibinfo {author} {\bibfnamefont {D.}~\bibnamefont
  {Walker}},\ }\bibfield  {title} {\bibinfo {title} {A review of thermal
  rectification observations and models in solid materials},\ }\href
  {https://doi.org/https://doi.org/10.1016/j.ijthermalsci.2010.12.004}
  {\bibfield  {journal} {\bibinfo  {journal} {Int. J. Therm. Sci.}\ }\textbf
  {\bibinfo {volume} {50}},\ \bibinfo {pages} {648 } (\bibinfo {year}
  {2011})}\BibitemShut {NoStop}%
\bibitem [{\citenamefont {Quan}\ \emph {et~al.}(2007)\citenamefont {Quan},
  \citenamefont {Liu}, \citenamefont {Sun},\ and\ \citenamefont
  {Nori}}]{PhysRevE.76.031105}%
  \BibitemOpen
  \bibfield  {author} {\bibinfo {author} {\bibfnamefont {H.~T.}\ \bibnamefont
  {Quan}}, \bibinfo {author} {\bibfnamefont {Y.-x.}\ \bibnamefont {Liu}},
  \bibinfo {author} {\bibfnamefont {C.~P.}\ \bibnamefont {Sun}},\ and\ \bibinfo
  {author} {\bibfnamefont {F.}~\bibnamefont {Nori}},\ }\bibfield  {title}
  {\bibinfo {title} {Quantum thermodynamic cycles and quantum heat engines},\
  }\href {https://doi.org/10.1103/PhysRevE.76.031105} {\bibfield  {journal}
  {\bibinfo  {journal} {Phys. Rev. E}\ }\textbf {\bibinfo {volume} {76}},\
  \bibinfo {pages} {031105} (\bibinfo {year} {2007})}\BibitemShut {NoStop}%
\bibitem [{\citenamefont {Ono}\ \emph {et~al.}(2020)\citenamefont {Ono},
  \citenamefont {Shevchenko}, \citenamefont {Mori}, \citenamefont {Moriyama},\
  and\ \citenamefont {Nori}}]{PhysRevLett.125.166802}%
  \BibitemOpen
  \bibfield  {author} {\bibinfo {author} {\bibfnamefont {K.}~\bibnamefont
  {Ono}}, \bibinfo {author} {\bibfnamefont {S.~N.}\ \bibnamefont {Shevchenko}},
  \bibinfo {author} {\bibfnamefont {T.}~\bibnamefont {Mori}}, \bibinfo {author}
  {\bibfnamefont {S.}~\bibnamefont {Moriyama}},\ and\ \bibinfo {author}
  {\bibfnamefont {F.}~\bibnamefont {Nori}},\ }\bibfield  {title} {\bibinfo
  {title} {Analog of a quantum heat engine using a single-spin qubit},\ }\href
  {https://doi.org/10.1103/PhysRevLett.125.166802} {\bibfield  {journal}
  {\bibinfo  {journal} {Phys. Rev. Lett.}\ }\textbf {\bibinfo {volume} {125}},\
  \bibinfo {pages} {166802} (\bibinfo {year} {2020})}\BibitemShut {NoStop}%
\bibitem [{\citenamefont {Cavaliere}\ \emph {et~al.}(2022)\citenamefont
  {Cavaliere}, \citenamefont {Carrega}, \citenamefont {De~Filippis},
  \citenamefont {Cataudella}, \citenamefont {Benenti},\ and\ \citenamefont
  {Sassetti}}]{cavaliere2022non}%
  \BibitemOpen
  \bibfield  {author} {\bibinfo {author} {\bibfnamefont {F.}~\bibnamefont
  {Cavaliere}}, \bibinfo {author} {\bibfnamefont {M.}~\bibnamefont {Carrega}},
  \bibinfo {author} {\bibfnamefont {G.}~\bibnamefont {De~Filippis}}, \bibinfo
  {author} {\bibfnamefont {V.}~\bibnamefont {Cataudella}}, \bibinfo {author}
  {\bibfnamefont {G.}~\bibnamefont {Benenti}},\ and\ \bibinfo {author}
  {\bibfnamefont {M.}~\bibnamefont {Sassetti}},\ }\bibfield  {title} {\bibinfo
  {title} {Dynamical heat engines with non-markovian reservoirs},\ }\href
  {https://doi.org/10.1103/PhysRevResearch.4.033233} {\bibfield  {journal}
  {\bibinfo  {journal} {Phys. Rev. Res.}\ }\textbf {\bibinfo {volume} {4}},\
  \bibinfo {pages} {033233} (\bibinfo {year} {2022})}\BibitemShut {NoStop}%
\bibitem [{\citenamefont {Bresque}\ \emph {et~al.}(2021)\citenamefont
  {Bresque}, \citenamefont {Camati}, \citenamefont {Rogers}, \citenamefont
  {Murch}, \citenamefont {Jordan},\ and\ \citenamefont
  {Auff\`eves}}]{PhysRevLett.126.120605}%
  \BibitemOpen
  \bibfield  {author} {\bibinfo {author} {\bibfnamefont {L.}~\bibnamefont
  {Bresque}}, \bibinfo {author} {\bibfnamefont {P.~A.}\ \bibnamefont {Camati}},
  \bibinfo {author} {\bibfnamefont {S.}~\bibnamefont {Rogers}}, \bibinfo
  {author} {\bibfnamefont {K.}~\bibnamefont {Murch}}, \bibinfo {author}
  {\bibfnamefont {A.~N.}\ \bibnamefont {Jordan}},\ and\ \bibinfo {author}
  {\bibfnamefont {A.}~\bibnamefont {Auff\`eves}},\ }\bibfield  {title}
  {\bibinfo {title} {Two-qubit engine fueled by entanglement and local
  measurements},\ }\href {https://doi.org/10.1103/PhysRevLett.126.120605}
  {\bibfield  {journal} {\bibinfo  {journal} {Phys. Rev. Lett.}\ }\textbf
  {\bibinfo {volume} {126}},\ \bibinfo {pages} {120605} (\bibinfo {year}
  {2021})}\BibitemShut {NoStop}%
\bibitem [{\citenamefont {Brask}\ \emph {et~al.}(2015)\citenamefont {Brask},
  \citenamefont {Haack}, \citenamefont {Brunner},\ and\ \citenamefont
  {Huber}}]{Bohr_Brask_2015}%
  \BibitemOpen
  \bibfield  {author} {\bibinfo {author} {\bibfnamefont {J.~B.}\ \bibnamefont
  {Brask}}, \bibinfo {author} {\bibfnamefont {G.}~\bibnamefont {Haack}},
  \bibinfo {author} {\bibfnamefont {N.}~\bibnamefont {Brunner}},\ and\ \bibinfo
  {author} {\bibfnamefont {M.}~\bibnamefont {Huber}},\ }\bibfield  {title}
  {\bibinfo {title} {Autonomous quantum thermal machine for generating
  steady-state entanglement},\ }\href
  {https://doi.org/10.1088/1367-2630/17/11/113029} {\bibfield  {journal}
  {\bibinfo  {journal} {New J. Phys.}\ }\textbf {\bibinfo {volume} {17}},\
  \bibinfo {pages} {113029} (\bibinfo {year} {2015})}\BibitemShut {NoStop}%
\bibitem [{\citenamefont {Lloyd}(1997)}]{PhysRevA.56.3374}%
  \BibitemOpen
  \bibfield  {author} {\bibinfo {author} {\bibfnamefont {S.}~\bibnamefont
  {Lloyd}},\ }\bibfield  {title} {\bibinfo {title} {Quantum-mechanical
  maxwell's demon},\ }\href {https://doi.org/10.1103/PhysRevA.56.3374}
  {\bibfield  {journal} {\bibinfo  {journal} {Phys. Rev. A}\ }\textbf {\bibinfo
  {volume} {56}},\ \bibinfo {pages} {3374} (\bibinfo {year}
  {1997})}\BibitemShut {NoStop}%
\bibitem [{\citenamefont {Elouard}\ \emph {et~al.}(2017)\citenamefont
  {Elouard}, \citenamefont {Herrera-Mart\'{\i}}, \citenamefont {Huard},\ and\
  \citenamefont {Auff\`eves}}]{PhysRevLett.118.260603}%
  \BibitemOpen
  \bibfield  {author} {\bibinfo {author} {\bibfnamefont {C.}~\bibnamefont
  {Elouard}}, \bibinfo {author} {\bibfnamefont {D.}~\bibnamefont
  {Herrera-Mart\'{\i}}}, \bibinfo {author} {\bibfnamefont {B.}~\bibnamefont
  {Huard}},\ and\ \bibinfo {author} {\bibfnamefont {A.}~\bibnamefont
  {Auff\`eves}},\ }\bibfield  {title} {\bibinfo {title} {Extracting work from
  quantum measurement in maxwell's demon engines},\ }\href
  {https://doi.org/10.1103/PhysRevLett.118.260603} {\bibfield  {journal}
  {\bibinfo  {journal} {Phys. Rev. Lett.}\ }\textbf {\bibinfo {volume} {118}},\
  \bibinfo {pages} {260603} (\bibinfo {year} {2017})}\BibitemShut {NoStop}%
\bibitem [{\citenamefont {Poulsen}\ \emph
  {et~al.}(2022{\natexlab{a}})\citenamefont {Poulsen}, \citenamefont {Majland},
  \citenamefont {Lloyd}, \citenamefont {Kjaergaard},\ and\ \citenamefont
  {Zinner}}]{PhysRevE.105.044141}%
  \BibitemOpen
  \bibfield  {author} {\bibinfo {author} {\bibfnamefont {K.}~\bibnamefont
  {Poulsen}}, \bibinfo {author} {\bibfnamefont {M.}~\bibnamefont {Majland}},
  \bibinfo {author} {\bibfnamefont {S.}~\bibnamefont {Lloyd}}, \bibinfo
  {author} {\bibfnamefont {M.}~\bibnamefont {Kjaergaard}},\ and\ \bibinfo
  {author} {\bibfnamefont {N.~T.}\ \bibnamefont {Zinner}},\ }\bibfield  {title}
  {\bibinfo {title} {Quantum maxwell's demon assisted by non-markovian
  effects},\ }\href {https://doi.org/10.1103/PhysRevE.105.044141} {\bibfield
  {journal} {\bibinfo  {journal} {Phys. Rev. E}\ }\textbf {\bibinfo {volume}
  {105}},\ \bibinfo {pages} {044141} (\bibinfo {year}
  {2022}{\natexlab{a}})}\BibitemShut {NoStop}%
\bibitem [{\citenamefont {Koski}\ \emph {et~al.}(2014)\citenamefont {Koski},
  \citenamefont {Maisi}, \citenamefont {Sagawa},\ and\ \citenamefont
  {Pekola}}]{PhysRevLett.113.030601}%
  \BibitemOpen
  \bibfield  {author} {\bibinfo {author} {\bibfnamefont {J.~V.}\ \bibnamefont
  {Koski}}, \bibinfo {author} {\bibfnamefont {V.~F.}\ \bibnamefont {Maisi}},
  \bibinfo {author} {\bibfnamefont {T.}~\bibnamefont {Sagawa}},\ and\ \bibinfo
  {author} {\bibfnamefont {J.~P.}\ \bibnamefont {Pekola}},\ }\bibfield  {title}
  {\bibinfo {title} {Experimental observation of the role of mutual information
  in the nonequilibrium dynamics of a {M}axwell demon},\ }\href
  {https://doi.org/10.1103/PhysRevLett.113.030601} {\bibfield  {journal}
  {\bibinfo  {journal} {Phys. Rev. Lett.}\ }\textbf {\bibinfo {volume} {113}},\
  \bibinfo {pages} {030601} (\bibinfo {year} {2014})}\BibitemShut {NoStop}%
\bibitem [{\citenamefont {Naghiloo}\ \emph {et~al.}(2018)\citenamefont
  {Naghiloo}, \citenamefont {Alonso}, \citenamefont {Romito}, \citenamefont
  {Lutz},\ and\ \citenamefont {Murch}}]{PhysRevLett.121.030604}%
  \BibitemOpen
  \bibfield  {author} {\bibinfo {author} {\bibfnamefont {M.}~\bibnamefont
  {Naghiloo}}, \bibinfo {author} {\bibfnamefont {J.~J.}\ \bibnamefont
  {Alonso}}, \bibinfo {author} {\bibfnamefont {A.}~\bibnamefont {Romito}},
  \bibinfo {author} {\bibfnamefont {E.}~\bibnamefont {Lutz}},\ and\ \bibinfo
  {author} {\bibfnamefont {K.~W.}\ \bibnamefont {Murch}},\ }\bibfield  {title}
  {\bibinfo {title} {Information gain and loss for a quantum {M}axwell's
  demon},\ }\href {https://doi.org/10.1103/PhysRevLett.121.030604} {\bibfield
  {journal} {\bibinfo  {journal} {Phys. Rev. Lett.}\ }\textbf {\bibinfo
  {volume} {121}},\ \bibinfo {pages} {030604} (\bibinfo {year}
  {2018})}\BibitemShut {NoStop}%
\bibitem [{\citenamefont {Senior}\ \emph {et~al.}(2020)\citenamefont {Senior},
  \citenamefont {Gubaydullin}, \citenamefont {Karimi}, \citenamefont
  {Peltonen}, \citenamefont {Ankerhold},\ and\ \citenamefont
  {Pekola}}]{Senior2020}%
  \BibitemOpen
  \bibfield  {author} {\bibinfo {author} {\bibfnamefont {J.}~\bibnamefont
  {Senior}}, \bibinfo {author} {\bibfnamefont {A.}~\bibnamefont {Gubaydullin}},
  \bibinfo {author} {\bibfnamefont {B.}~\bibnamefont {Karimi}}, \bibinfo
  {author} {\bibfnamefont {J.~T.}\ \bibnamefont {Peltonen}}, \bibinfo {author}
  {\bibfnamefont {J.}~\bibnamefont {Ankerhold}},\ and\ \bibinfo {author}
  {\bibfnamefont {J.~P.}\ \bibnamefont {Pekola}},\ }\bibfield  {title}
  {\bibinfo {title} {Heat rectification via a superconducting artificial
  atom},\ }\href {https://doi.org/10.1038/s42005-020-0307-5} {\bibfield
  {journal} {\bibinfo  {journal} {Commun. Phys.}\ }\textbf {\bibinfo {volume}
  {3}},\ \bibinfo {pages} {40} (\bibinfo {year} {2020})}\BibitemShut {NoStop}%
\bibitem [{\citenamefont {Ronzani}\ \emph {et~al.}(2018)\citenamefont
  {Ronzani}, \citenamefont {Karimi}, \citenamefont {Senior}, \citenamefont
  {Chang}, \citenamefont {Peltonen}, \citenamefont {Chen},\ and\ \citenamefont
  {Pekola}}]{Ronzani2018}%
  \BibitemOpen
  \bibfield  {author} {\bibinfo {author} {\bibfnamefont {A.}~\bibnamefont
  {Ronzani}}, \bibinfo {author} {\bibfnamefont {B.}~\bibnamefont {Karimi}},
  \bibinfo {author} {\bibfnamefont {J.}~\bibnamefont {Senior}}, \bibinfo
  {author} {\bibfnamefont {Y.-C.}\ \bibnamefont {Chang}}, \bibinfo {author}
  {\bibfnamefont {J.~T.}\ \bibnamefont {Peltonen}}, \bibinfo {author}
  {\bibfnamefont {C.}~\bibnamefont {Chen}},\ and\ \bibinfo {author}
  {\bibfnamefont {J.~P.}\ \bibnamefont {Pekola}},\ }\bibfield  {title}
  {\bibinfo {title} {Tunable photonic heat transport in a quantum heat valve},\
  }\href {https://doi.org/10.1038/s41567-018-0199-4} {\bibfield  {journal}
  {\bibinfo  {journal} {Nat. Phys.}\ }\textbf {\bibinfo {volume} {14}},\
  \bibinfo {pages} {991} (\bibinfo {year} {2018})}\BibitemShut {NoStop}%
\bibitem [{\citenamefont {Gubaydullin}\ \emph {et~al.}(2022)\citenamefont
  {Gubaydullin}, \citenamefont {Thomas}, \citenamefont {Golubev}, \citenamefont
  {Lvov}, \citenamefont {Peltonen},\ and\ \citenamefont
  {Pekola}}]{Gubaydullin2022}%
  \BibitemOpen
  \bibfield  {author} {\bibinfo {author} {\bibfnamefont {A.}~\bibnamefont
  {Gubaydullin}}, \bibinfo {author} {\bibfnamefont {G.}~\bibnamefont {Thomas}},
  \bibinfo {author} {\bibfnamefont {D.~S.}\ \bibnamefont {Golubev}}, \bibinfo
  {author} {\bibfnamefont {D.}~\bibnamefont {Lvov}}, \bibinfo {author}
  {\bibfnamefont {J.~T.}\ \bibnamefont {Peltonen}},\ and\ \bibinfo {author}
  {\bibfnamefont {J.~P.}\ \bibnamefont {Pekola}},\ }\bibfield  {title}
  {\bibinfo {title} {Photonic heat transport in three terminal superconducting
  circuit},\ }\href {https://doi.org/10.1038/s41467-022-29078-x} {\bibfield
  {journal} {\bibinfo  {journal} {Nat. Commun.}\ }\textbf {\bibinfo {volume}
  {13}},\ \bibinfo {pages} {1552} (\bibinfo {year} {2022})}\BibitemShut
  {NoStop}%
\bibitem [{\citenamefont {Joulain}\ \emph {et~al.}(2016)\citenamefont
  {Joulain}, \citenamefont {Drevillon}, \citenamefont {Ezzahri},\ and\
  \citenamefont {Ordonez-Miranda}}]{PhysRevLett.116.200601}%
  \BibitemOpen
  \bibfield  {author} {\bibinfo {author} {\bibfnamefont {K.}~\bibnamefont
  {Joulain}}, \bibinfo {author} {\bibfnamefont {J.}~\bibnamefont {Drevillon}},
  \bibinfo {author} {\bibfnamefont {Y.}~\bibnamefont {Ezzahri}},\ and\ \bibinfo
  {author} {\bibfnamefont {J.}~\bibnamefont {Ordonez-Miranda}},\ }\bibfield
  {title} {\bibinfo {title} {Quantum thermal transistor},\ }\href
  {https://doi.org/10.1103/PhysRevLett.116.200601} {\bibfield  {journal}
  {\bibinfo  {journal} {Phys. Rev. Lett.}\ }\textbf {\bibinfo {volume} {116}},\
  \bibinfo {pages} {200601} (\bibinfo {year} {2016})}\BibitemShut {NoStop}%
\bibitem [{\citenamefont {Majland}\ \emph {et~al.}(2020)\citenamefont
  {Majland}, \citenamefont {Christensen},\ and\ \citenamefont
  {Zinner}}]{PhysRevB.101.184510}%
  \BibitemOpen
  \bibfield  {author} {\bibinfo {author} {\bibfnamefont {M.}~\bibnamefont
  {Majland}}, \bibinfo {author} {\bibfnamefont {K.~S.}\ \bibnamefont
  {Christensen}},\ and\ \bibinfo {author} {\bibfnamefont {N.~T.}\ \bibnamefont
  {Zinner}},\ }\bibfield  {title} {\bibinfo {title} {Quantum thermal transistor
  in superconducting circuits},\ }\href
  {https://doi.org/10.1103/PhysRevB.101.184510} {\bibfield  {journal} {\bibinfo
   {journal} {Phys. Rev. B}\ }\textbf {\bibinfo {volume} {101}},\ \bibinfo
  {pages} {184510} (\bibinfo {year} {2020})}\BibitemShut {NoStop}%
\bibitem [{\citenamefont {Guo}\ \emph {et~al.}(2018)\citenamefont {Guo},
  \citenamefont {Liu},\ and\ \citenamefont {Yu}}]{PhysRevE.98.022118}%
  \BibitemOpen
  \bibfield  {author} {\bibinfo {author} {\bibfnamefont {B.-q.}\ \bibnamefont
  {Guo}}, \bibinfo {author} {\bibfnamefont {T.}~\bibnamefont {Liu}},\ and\
  \bibinfo {author} {\bibfnamefont {C.-s.}\ \bibnamefont {Yu}},\ }\bibfield
  {title} {\bibinfo {title} {Quantum thermal transistor based on qubit-qutrit
  coupling},\ }\href {https://doi.org/10.1103/PhysRevE.98.022118} {\bibfield
  {journal} {\bibinfo  {journal} {Phys. Rev. E}\ }\textbf {\bibinfo {volume}
  {98}},\ \bibinfo {pages} {022118} (\bibinfo {year} {2018})}\BibitemShut
  {NoStop}%
\bibitem [{\citenamefont {Poulsen}\ and\ \citenamefont
  {Zinner}(2022)}]{poulsen2022dark}%
  \BibitemOpen
  \bibfield  {author} {\bibinfo {author} {\bibfnamefont {K.}~\bibnamefont
  {Poulsen}}\ and\ \bibinfo {author} {\bibfnamefont {N.~T.}\ \bibnamefont
  {Zinner}},\ }\bibfield  {title} {\bibinfo {title} {Dark-state-induced heat
  rectification},\ }\href {https://doi.org/10.1103/PhysRevE.106.034116}
  {\bibfield  {journal} {\bibinfo  {journal} {Phys. Rev. E}\ }\textbf {\bibinfo
  {volume} {106}},\ \bibinfo {pages} {034116} (\bibinfo {year}
  {2022})}\BibitemShut {NoStop}%
\bibitem [{\citenamefont {Marcos-Vicioso}\ \emph {et~al.}(2018)\citenamefont
  {Marcos-Vicioso}, \citenamefont {L\'opez-Jurado}, \citenamefont
  {Ruiz-Garcia},\ and\ \citenamefont {S\'anchez}}]{PhysRevB.98.035414}%
  \BibitemOpen
  \bibfield  {author} {\bibinfo {author} {\bibfnamefont {A.}~\bibnamefont
  {Marcos-Vicioso}}, \bibinfo {author} {\bibfnamefont {C.}~\bibnamefont
  {L\'opez-Jurado}}, \bibinfo {author} {\bibfnamefont {M.}~\bibnamefont
  {Ruiz-Garcia}},\ and\ \bibinfo {author} {\bibfnamefont {R.}~\bibnamefont
  {S\'anchez}},\ }\bibfield  {title} {\bibinfo {title} {Thermal rectification
  with interacting electronic channels: Exploiting degeneracy, quantum
  superpositions, and interference},\ }\href
  {https://doi.org/10.1103/PhysRevB.98.035414} {\bibfield  {journal} {\bibinfo
  {journal} {Phys. Rev. B}\ }\textbf {\bibinfo {volume} {98}},\ \bibinfo
  {pages} {035414} (\bibinfo {year} {2018})}\BibitemShut {NoStop}%
\bibitem [{\citenamefont {D{\'{\i}}az}\ and\ \citenamefont
  {S{\'{a}}nchez}(2021)}]{D_az_2021}%
  \BibitemOpen
  \bibfield  {author} {\bibinfo {author} {\bibfnamefont {I.}~\bibnamefont
  {D{\'{\i}}az}}\ and\ \bibinfo {author} {\bibfnamefont {R.}~\bibnamefont
  {S{\'{a}}nchez}},\ }\bibfield  {title} {\bibinfo {title} {The qutrit as a
  heat diode and circulator},\ }\href
  {https://doi.org/10.1088/1367-2630/ac4211} {\bibfield  {journal} {\bibinfo
  {journal} {New J. Phys.}\ }\textbf {\bibinfo {volume} {23}},\ \bibinfo
  {pages} {125006} (\bibinfo {year} {2021})}\BibitemShut {NoStop}%
\bibitem [{\citenamefont {Iorio}\ \emph {et~al.}(2021)\citenamefont {Iorio},
  \citenamefont {Strambini}, \citenamefont {Haack}, \citenamefont {Campisi},\
  and\ \citenamefont {Giazotto}}]{PhysRevApplied.15.054050}%
  \BibitemOpen
  \bibfield  {author} {\bibinfo {author} {\bibfnamefont {A.}~\bibnamefont
  {Iorio}}, \bibinfo {author} {\bibfnamefont {E.}~\bibnamefont {Strambini}},
  \bibinfo {author} {\bibfnamefont {G.}~\bibnamefont {Haack}}, \bibinfo
  {author} {\bibfnamefont {M.}~\bibnamefont {Campisi}},\ and\ \bibinfo {author}
  {\bibfnamefont {F.}~\bibnamefont {Giazotto}},\ }\bibfield  {title} {\bibinfo
  {title} {Photonic heat rectification in a system of coupled qubits},\ }\href
  {https://doi.org/10.1103/PhysRevApplied.15.054050} {\bibfield  {journal}
  {\bibinfo  {journal} {Phys. Rev. Appl.}\ }\textbf {\bibinfo {volume} {15}},\
  \bibinfo {pages} {054050} (\bibinfo {year} {2021})}\BibitemShut {NoStop}%
\bibitem [{\citenamefont {Werlang}\ \emph {et~al.}(2014)\citenamefont
  {Werlang}, \citenamefont {Marchiori}, \citenamefont {Cornelio},\ and\
  \citenamefont {Valente}}]{PhysRevE.89.062109}%
  \BibitemOpen
  \bibfield  {author} {\bibinfo {author} {\bibfnamefont {T.}~\bibnamefont
  {Werlang}}, \bibinfo {author} {\bibfnamefont {M.~A.}\ \bibnamefont
  {Marchiori}}, \bibinfo {author} {\bibfnamefont {M.~F.}\ \bibnamefont
  {Cornelio}},\ and\ \bibinfo {author} {\bibfnamefont {D.}~\bibnamefont
  {Valente}},\ }\bibfield  {title} {\bibinfo {title} {Optimal rectification in
  the ultrastrong coupling regime},\ }\href
  {https://doi.org/10.1103/PhysRevE.89.062109} {\bibfield  {journal} {\bibinfo
  {journal} {Phys. Rev. E}\ }\textbf {\bibinfo {volume} {89}},\ \bibinfo
  {pages} {062109} (\bibinfo {year} {2014})}\BibitemShut {NoStop}%
\bibitem [{\citenamefont {Segal}(2006)}]{PhysRevB.73.205415}%
  \BibitemOpen
  \bibfield  {author} {\bibinfo {author} {\bibfnamefont {D.}~\bibnamefont
  {Segal}},\ }\bibfield  {title} {\bibinfo {title} {Heat flow in nonlinear
  molecular junctions: Master equation analysis},\ }\href
  {https://doi.org/10.1103/PhysRevB.73.205415} {\bibfield  {journal} {\bibinfo
  {journal} {Phys. Rev. B}\ }\textbf {\bibinfo {volume} {73}},\ \bibinfo
  {pages} {205415} (\bibinfo {year} {2006})}\BibitemShut {NoStop}%
\bibitem [{\citenamefont {Carrega}\ \emph {et~al.}(2022)\citenamefont
  {Carrega}, \citenamefont {Cangemi}, \citenamefont {De~Filippis},
  \citenamefont {Cataudella}, \citenamefont {Benenti},\ and\ \citenamefont
  {Sassetti}}]{PRXQuantum.3.010323}%
  \BibitemOpen
  \bibfield  {author} {\bibinfo {author} {\bibfnamefont {M.}~\bibnamefont
  {Carrega}}, \bibinfo {author} {\bibfnamefont {L.~M.}\ \bibnamefont
  {Cangemi}}, \bibinfo {author} {\bibfnamefont {G.}~\bibnamefont
  {De~Filippis}}, \bibinfo {author} {\bibfnamefont {V.}~\bibnamefont
  {Cataudella}}, \bibinfo {author} {\bibfnamefont {G.}~\bibnamefont
  {Benenti}},\ and\ \bibinfo {author} {\bibfnamefont {M.}~\bibnamefont
  {Sassetti}},\ }\bibfield  {title} {\bibinfo {title} {Engineering dynamical
  couplings for quantum thermodynamic tasks},\ }\href
  {https://doi.org/10.1103/PRXQuantum.3.010323} {\bibfield  {journal} {\bibinfo
   {journal} {PRX Quantum}\ }\textbf {\bibinfo {volume} {3}},\ \bibinfo {pages}
  {010323} (\bibinfo {year} {2022})}\BibitemShut {NoStop}%
\bibitem [{\citenamefont {Bhandari}\ \emph {et~al.}(2021)\citenamefont
  {Bhandari}, \citenamefont {Erdman}, \citenamefont {Fazio}, \citenamefont
  {Paladino},\ and\ \citenamefont {Taddei}}]{PhysRevB.103.155434}%
  \BibitemOpen
  \bibfield  {author} {\bibinfo {author} {\bibfnamefont {B.}~\bibnamefont
  {Bhandari}}, \bibinfo {author} {\bibfnamefont {P.~A.}\ \bibnamefont
  {Erdman}}, \bibinfo {author} {\bibfnamefont {R.}~\bibnamefont {Fazio}},
  \bibinfo {author} {\bibfnamefont {E.}~\bibnamefont {Paladino}},\ and\
  \bibinfo {author} {\bibfnamefont {F.}~\bibnamefont {Taddei}},\ }\bibfield
  {title} {\bibinfo {title} {Thermal rectification through a nonlinear quantum
  resonator},\ }\href {https://doi.org/10.1103/PhysRevB.103.155434} {\bibfield
  {journal} {\bibinfo  {journal} {Phys. Rev. B}\ }\textbf {\bibinfo {volume}
  {103}},\ \bibinfo {pages} {155434} (\bibinfo {year} {2021})}\BibitemShut
  {NoStop}%
\bibitem [{\citenamefont {Tesser}\ \emph {et~al.}(2022)\citenamefont {Tesser},
  \citenamefont {Bhandari}, \citenamefont {Erdman}, \citenamefont {Paladino},
  \citenamefont {Fazio},\ and\ \citenamefont {Taddei}}]{Tesser_2022}%
  \BibitemOpen
  \bibfield  {author} {\bibinfo {author} {\bibfnamefont {L.}~\bibnamefont
  {Tesser}}, \bibinfo {author} {\bibfnamefont {B.}~\bibnamefont {Bhandari}},
  \bibinfo {author} {\bibfnamefont {P.~A.}\ \bibnamefont {Erdman}}, \bibinfo
  {author} {\bibfnamefont {E.}~\bibnamefont {Paladino}}, \bibinfo {author}
  {\bibfnamefont {R.}~\bibnamefont {Fazio}},\ and\ \bibinfo {author}
  {\bibfnamefont {F.}~\bibnamefont {Taddei}},\ }\bibfield  {title} {\bibinfo
  {title} {Heat rectification through single and coupled quantum dots},\ }\href
  {https://doi.org/10.1088/1367-2630/ac53b8} {\bibfield  {journal} {\bibinfo
  {journal} {New J. Phys.}\ }\textbf {\bibinfo {volume} {24}},\ \bibinfo
  {pages} {035001} (\bibinfo {year} {2022})}\BibitemShut {NoStop}%
\bibitem [{\citenamefont {Poulsen}\ \emph
  {et~al.}(2022{\natexlab{b}})\citenamefont {Poulsen}, \citenamefont {Santos},
  \citenamefont {Kristensen},\ and\ \citenamefont
  {Zinner}}]{PhysRevA.105.052605}%
  \BibitemOpen
  \bibfield  {author} {\bibinfo {author} {\bibfnamefont {K.}~\bibnamefont
  {Poulsen}}, \bibinfo {author} {\bibfnamefont {A.~C.}\ \bibnamefont {Santos}},
  \bibinfo {author} {\bibfnamefont {L.~B.}\ \bibnamefont {Kristensen}},\ and\
  \bibinfo {author} {\bibfnamefont {N.~T.}\ \bibnamefont {Zinner}},\ }\bibfield
   {title} {\bibinfo {title} {Entanglement-enhanced quantum rectification},\
  }\href {https://doi.org/10.1103/PhysRevA.105.052605} {\bibfield  {journal}
  {\bibinfo  {journal} {Phys. Rev. A}\ }\textbf {\bibinfo {volume} {105}},\
  \bibinfo {pages} {052605} (\bibinfo {year} {2022}{\natexlab{b}})}\BibitemShut
  {NoStop}%
\bibitem [{\citenamefont {Pollak}(1886)}]{patent}%
  \BibitemOpen
  \bibfield  {author} {\bibinfo {author} {\bibfnamefont {K.}~\bibnamefont
  {Pollak}},\ }\href@noop {} {\bibinfo {title} {Improvements in means for
  controlling or directing electric currents.}} (\bibinfo {year} {British
  patent 24398, Dec. 1886})\BibitemShut {NoStop}%
\bibitem [{\citenamefont {Lindblad}(1976)}]{Lindblad1976}%
  \BibitemOpen
  \bibfield  {author} {\bibinfo {author} {\bibfnamefont {G.}~\bibnamefont
  {Lindblad}},\ }\bibfield  {title} {\bibinfo {title} {On the generators of
  quantum dynamical semigroups},\ }\href {https://doi.org/10.1007/BF01608499}
  {\bibfield  {journal} {\bibinfo  {journal} {Commun. Math. Phys.}\ }\textbf
  {\bibinfo {volume} {48}},\ \bibinfo {pages} {119} (\bibinfo {year}
  {1976})}\BibitemShut {NoStop}%
\bibitem [{\citenamefont {Breuer}\ and\ \citenamefont
  {Petruccione}(2002)}]{breuer2002theory}%
  \BibitemOpen
  \bibfield  {author} {\bibinfo {author} {\bibfnamefont {H.}~\bibnamefont
  {Breuer}}\ and\ \bibinfo {author} {\bibfnamefont {F.}~\bibnamefont
  {Petruccione}},\ }\href@noop {} {\emph {\bibinfo {title} {The theory of open
  quantum systems}}}\ (\bibinfo  {publisher} {Oxford University Press,
  Oxford},\ \bibinfo {year} {2002})\BibitemShut {NoStop}%
\bibitem [{\citenamefont {Poulsen}\ \emph
  {et~al.}(2022{\natexlab{c}})\citenamefont {Poulsen}, \citenamefont {Santos},\
  and\ \citenamefont {Zinner}}]{poulsen2021quantum}%
  \BibitemOpen
  \bibfield  {author} {\bibinfo {author} {\bibfnamefont {K.}~\bibnamefont
  {Poulsen}}, \bibinfo {author} {\bibfnamefont {A.~C.}\ \bibnamefont
  {Santos}},\ and\ \bibinfo {author} {\bibfnamefont {N.~T.}\ \bibnamefont
  {Zinner}},\ }\bibfield  {title} {\bibinfo {title} {Quantum wheatstone
  bridge},\ }\href {https://doi.org/10.1103/PhysRevLett.128.240401} {\bibfield
  {journal} {\bibinfo  {journal} {Phys. Rev. Lett.}\ }\textbf {\bibinfo
  {volume} {128}},\ \bibinfo {pages} {240401} (\bibinfo {year}
  {2022}{\natexlab{c}})}\BibitemShut {NoStop}%
\bibitem [{\citenamefont {Kim}\ \emph {et~al.}(2023)\citenamefont {Kim},
  \citenamefont {Eddins}, \citenamefont {Anand}, \citenamefont {Wei},
  \citenamefont {van~den Berg}, \citenamefont {Rosenblatt}, \citenamefont
  {Nayfeh}, \citenamefont {Wu}, \citenamefont {Zaletel}, \citenamefont
  {Temme},\ and\ \citenamefont {Kandala}}]{Kim2023}%
  \BibitemOpen
  \bibfield  {author} {\bibinfo {author} {\bibfnamefont {Y.}~\bibnamefont
  {Kim}}, \bibinfo {author} {\bibfnamefont {A.}~\bibnamefont {Eddins}},
  \bibinfo {author} {\bibfnamefont {S.}~\bibnamefont {Anand}}, \bibinfo
  {author} {\bibfnamefont {K.~X.}\ \bibnamefont {Wei}}, \bibinfo {author}
  {\bibfnamefont {E.}~\bibnamefont {van~den Berg}}, \bibinfo {author}
  {\bibfnamefont {S.}~\bibnamefont {Rosenblatt}}, \bibinfo {author}
  {\bibfnamefont {H.}~\bibnamefont {Nayfeh}}, \bibinfo {author} {\bibfnamefont
  {Y.}~\bibnamefont {Wu}}, \bibinfo {author} {\bibfnamefont {M.}~\bibnamefont
  {Zaletel}}, \bibinfo {author} {\bibfnamefont {K.}~\bibnamefont {Temme}},\
  and\ \bibinfo {author} {\bibfnamefont {A.}~\bibnamefont {Kandala}},\
  }\bibfield  {title} {\bibinfo {title} {Evidence for the utility of quantum
  computing before fault tolerance},\ }\href
  {https://doi.org/10.1038/s41586-023-06096-3} {\bibfield  {journal} {\bibinfo
  {journal} {Nature}\ }\textbf {\bibinfo {volume} {618}},\ \bibinfo {pages}
  {500} (\bibinfo {year} {2023})}\BibitemShut {NoStop}%
\bibitem [{\citenamefont {Kjaergaard}\ \emph {et~al.}(2020)\citenamefont
  {Kjaergaard}, \citenamefont {Schwartz}, \citenamefont {Braumüller},
  \citenamefont {Krantz}, \citenamefont {Wang}, \citenamefont {Gustavsson},\
  and\ \citenamefont {Oliver}}]{doi:10.1146/annurev-conmatphys-031119-050605}%
  \BibitemOpen
  \bibfield  {author} {\bibinfo {author} {\bibfnamefont {M.}~\bibnamefont
  {Kjaergaard}}, \bibinfo {author} {\bibfnamefont {M.~E.}\ \bibnamefont
  {Schwartz}}, \bibinfo {author} {\bibfnamefont {J.}~\bibnamefont
  {Braumüller}}, \bibinfo {author} {\bibfnamefont {P.}~\bibnamefont {Krantz}},
  \bibinfo {author} {\bibfnamefont {J.~I.-J.}\ \bibnamefont {Wang}}, \bibinfo
  {author} {\bibfnamefont {S.}~\bibnamefont {Gustavsson}},\ and\ \bibinfo
  {author} {\bibfnamefont {W.~D.}\ \bibnamefont {Oliver}},\ }\bibfield  {title}
  {\bibinfo {title} {Superconducting qubits: Current state of play},\ }\href
  {https://doi.org/10.1146/annurev-conmatphys-031119-050605} {\bibfield
  {journal} {\bibinfo  {journal} {Annu. Rev. Condens. Matter Phys.}\ }\textbf
  {\bibinfo {volume} {11}},\ \bibinfo {pages} {369} (\bibinfo {year}
  {2020})}\BibitemShut {NoStop}%
\bibitem [{\citenamefont {Yan}\ \emph {et~al.}(2018)\citenamefont {Yan},
  \citenamefont {Krantz}, \citenamefont {Sung}, \citenamefont {Kjaergaard},
  \citenamefont {Campbell}, \citenamefont {Orlando}, \citenamefont
  {Gustavsson},\ and\ \citenamefont {Oliver}}]{PhysRevApplied.10.054062}%
  \BibitemOpen
  \bibfield  {author} {\bibinfo {author} {\bibfnamefont {F.}~\bibnamefont
  {Yan}}, \bibinfo {author} {\bibfnamefont {P.}~\bibnamefont {Krantz}},
  \bibinfo {author} {\bibfnamefont {Y.}~\bibnamefont {Sung}}, \bibinfo {author}
  {\bibfnamefont {M.}~\bibnamefont {Kjaergaard}}, \bibinfo {author}
  {\bibfnamefont {D.~L.}\ \bibnamefont {Campbell}}, \bibinfo {author}
  {\bibfnamefont {T.~P.}\ \bibnamefont {Orlando}}, \bibinfo {author}
  {\bibfnamefont {S.}~\bibnamefont {Gustavsson}},\ and\ \bibinfo {author}
  {\bibfnamefont {W.~D.}\ \bibnamefont {Oliver}},\ }\bibfield  {title}
  {\bibinfo {title} {Tunable coupling scheme for implementing high-fidelity
  two-qubit gates},\ }\href {https://doi.org/10.1103/PhysRevApplied.10.054062}
  {\bibfield  {journal} {\bibinfo  {journal} {Phys. Rev. Appl.}\ }\textbf
  {\bibinfo {volume} {10}},\ \bibinfo {pages} {054062} (\bibinfo {year}
  {2018})}\BibitemShut {NoStop}%
\bibitem [{\citenamefont {Johanning}\ \emph {et~al.}(2009)\citenamefont
  {Johanning}, \citenamefont {Var{\'o}n},\ and\ \citenamefont
  {Wunderlich}}]{Johanning_2009}%
  \BibitemOpen
  \bibfield  {author} {\bibinfo {author} {\bibfnamefont {M.}~\bibnamefont
  {Johanning}}, \bibinfo {author} {\bibfnamefont {A.~F.}\ \bibnamefont
  {Var{\'o}n}},\ and\ \bibinfo {author} {\bibfnamefont {C.}~\bibnamefont
  {Wunderlich}},\ }\bibfield  {title} {\bibinfo {title} {Quantum simulations
  with cold trapped ions},\ }\href
  {https://doi.org/10.1088/0953-4075/42/15/154009} {\bibfield  {journal}
  {\bibinfo  {journal} {J. Phys. B}\ }\textbf {\bibinfo {volume} {42}},\
  \bibinfo {pages} {154009} (\bibinfo {year} {2009})}\BibitemShut {NoStop}%
\bibitem [{\citenamefont {Häffner}\ \emph {et~al.}(2008)\citenamefont
  {Häffner}, \citenamefont {Roos},\ and\ \citenamefont
  {Blatt}}]{HAFFNER2008155}%
  \BibitemOpen
  \bibfield  {author} {\bibinfo {author} {\bibfnamefont {H.}~\bibnamefont
  {Häffner}}, \bibinfo {author} {\bibfnamefont {C.}~\bibnamefont {Roos}},\
  and\ \bibinfo {author} {\bibfnamefont {R.}~\bibnamefont {Blatt}},\ }\bibfield
   {title} {\bibinfo {title} {Quantum computing with trapped ions},\ }\href
  {https://doi.org/https://doi.org/10.1016/j.physrep.2008.09.003} {\bibfield
  {journal} {\bibinfo  {journal} {Phys. Rep.}\ }\textbf {\bibinfo {volume}
  {469}},\ \bibinfo {pages} {155} (\bibinfo {year} {2008})}\BibitemShut
  {NoStop}%
\bibitem [{\citenamefont {Dann}\ \emph {et~al.}(2018)\citenamefont {Dann},
  \citenamefont {Levy},\ and\ \citenamefont {Kosloff}}]{PhysRevA.98.052129}%
  \BibitemOpen
  \bibfield  {author} {\bibinfo {author} {\bibfnamefont {R.}~\bibnamefont
  {Dann}}, \bibinfo {author} {\bibfnamefont {A.}~\bibnamefont {Levy}},\ and\
  \bibinfo {author} {\bibfnamefont {R.}~\bibnamefont {Kosloff}},\ }\bibfield
  {title} {\bibinfo {title} {Time-dependent markovian quantum master
  equation},\ }\href {https://doi.org/10.1103/PhysRevA.98.052129} {\bibfield
  {journal} {\bibinfo  {journal} {Phys. Rev. A}\ }\textbf {\bibinfo {volume}
  {98}},\ \bibinfo {pages} {052129} (\bibinfo {year} {2018})}\BibitemShut
  {NoStop}%
\bibitem [{\citenamefont {Levy}\ \emph {et~al.}(2012)\citenamefont {Levy},
  \citenamefont {Alicki},\ and\ \citenamefont {Kosloff}}]{PhysRevE.85.061126}%
  \BibitemOpen
  \bibfield  {author} {\bibinfo {author} {\bibfnamefont {A.}~\bibnamefont
  {Levy}}, \bibinfo {author} {\bibfnamefont {R.}~\bibnamefont {Alicki}},\ and\
  \bibinfo {author} {\bibfnamefont {R.}~\bibnamefont {Kosloff}},\ }\bibfield
  {title} {\bibinfo {title} {Quantum refrigerators and the third law of
  thermodynamics},\ }\href {https://doi.org/10.1103/PhysRevE.85.061126}
  {\bibfield  {journal} {\bibinfo  {journal} {Phys. Rev. E}\ }\textbf {\bibinfo
  {volume} {85}},\ \bibinfo {pages} {061126} (\bibinfo {year}
  {2012})}\BibitemShut {NoStop}%
\bibitem [{\citenamefont {Alicki}\ \emph {et~al.}(2006)\citenamefont {Alicki},
  \citenamefont {Lidar},\ and\ \citenamefont {Zanardi}}]{PhysRevA.73.052311}%
  \BibitemOpen
  \bibfield  {author} {\bibinfo {author} {\bibfnamefont {R.}~\bibnamefont
  {Alicki}}, \bibinfo {author} {\bibfnamefont {D.~A.}\ \bibnamefont {Lidar}},\
  and\ \bibinfo {author} {\bibfnamefont {P.}~\bibnamefont {Zanardi}},\
  }\bibfield  {title} {\bibinfo {title} {Internal consistency of fault-tolerant
  quantum error correction in light of rigorous derivations of the quantum
  markovian limit},\ }\href {https://doi.org/10.1103/PhysRevA.73.052311}
  {\bibfield  {journal} {\bibinfo  {journal} {Phys. Rev. A}\ }\textbf {\bibinfo
  {volume} {73}},\ \bibinfo {pages} {052311} (\bibinfo {year}
  {2006})}\BibitemShut {NoStop}%
\end{thebibliography}
\end{document}